\documentclass[10pt]{article}
\usepackage{graphicx}
\usepackage{amsmath}
\usepackage{hanging}
\usepackage{apacite}
\usepackage[lmargin=0.5in, rmargin=0.5in, tmargin=1in, bmargin=1in]{geometry}
\usepackage[labelfont=bf]{caption}
\usepackage{authblk}

\usepackage{caption} 
\usepackage{lettrine}
\usepackage{kpfonts}
\usepackage{multicol}
\usepackage{titling}
\usepackage{color}
\usepackage{dblfloatfix}
\usepackage{etex}
\usepackage[figuresright]{rotating}
\usepackage{amssymb}
\usepackage{bbm}
\usepackage{subfigure}
\usepackage{pstricks,pst-node,pst-plot}
\usepackage{color}
\usepackage{pst-all}
\usepackage{pspicture}
\usepackage{booktabs}
\usepackage{wasysym}
\usepackage[all]{xy}
\usepackage{epsfig,graphicx}
\usepackage{theorem}
\usepackage{tikz}
\usepackage{algorithm}
\usepackage{algorithmic}
\usepackage{multirow}
\usepackage[flushleft]{threeparttable}
\makeatletter

\definecolor{blue}{rgb}{0,0,0.5}

 \pretitle{\begin{flushleft}\LARGE}
 \posttitle{\end{flushleft}}
  \preauthor{\begin{flushleft}}
 \postauthor{\end{flushleft}}
  
  \usepackage{titlesec}
\titlespacing*{\section}{0pt}{10pt}{-1pt}
 \titlespacing*{\subsection}{0pt}{2pt}{-1pt}
 \titlespacing*{\subsubsection}{0pt}{2pt}{-1pt}
\titleformat*{\section}{\normalfont\Large\bfseries\color{blue}}
\titleformat*{\subsection}{\normalfont\large\bfseries\color{blue}}
\titleformat*{\subsubsection}{\normalfont\bfseries\color{blue}}

\newcommand{\Prob}{\mathbb{P}}

\newcounter{count}
\newcounter{asscount}

\makeatletter 
\renewcommand\small{\@setfontsize\small{8.8pt}{18}}
\makeatother

\bibleftmargin{10pt}
\bibindent{-10pt}

\title{\vspace{-2.0cm}\textbf{\textcolor{blue}{Network Structure Explains the Impact of Attitudes on Voting Decisions}}}
\date{}
\author{Jonas Dalege, Denny Borsboom, Frenk van Harreveld, Lourens J. Waldorp \& Han L. J. van der Maas}
\affil{Department of Psychology, University of Amsterdam, 1018 WT Amsterdam, The Netherlands}

\begin{document}

\maketitle

\section*{\vspace{-8ex}} 
\textbf{Attitudes can have a profound impact on socially relevant behaviours, such as voting. However, this effect is not uniform across situations or individuals, and it is at present difficult to predict whether attitudes will predict behaviour in any given circumstance. Using a network model, we demonstrate that (a) more strongly connected attitude networks have a stronger impact on behaviour, and (b) within any given attitude network, the most central attitude elements have the strongest impact. We test these hypotheses using data on voting and attitudes toward presidential candidates in the US presidential elections from 1980 to 2012. These analyses confirm that the predictive value of attitude networks depends almost entirely on their level of connectivity, with more central attitude elements having stronger impact. The impact of attitudes on voting behaviour can thus be reliably determined before elections take place by using network analyses.}

\begin{multicols}{2}

\lettrine[findent=0pt, lines=3]{\textcolor{blue}S}{ }uppose you are one of the more than 130 million Americans who voted in the presidential election in 2016. Let us further assume that you were supportive of Hillary Clinton: You mostly held positive beliefs (e.g., you thought she was a good leader and a knowledgeable person) and you had positive feelings toward her (e.g., she made you feel hopeful and proud), representing a positive attitude toward Hillary Clinton \cite{Dalege2016, Eagly1993, Fishbein1975, Rosenberg1960}. However, you also held a few negative beliefs toward her (e.g., you thought that Hillary Clinton was not very honest). Did your overall positive attitude cause you to vote for Hillary Clinton? Here we show that the answer to this question depends on the network structure of your attitude: First, we show that the impact of attitudes (i.e., average of the attitude elements) on behavioural decisions depends on the connectivity of the attitude network (e.g., your the network of your positive attitude toward Hillary Clinton was highly connected, so you probably voted for Hillary Clinton). Second, we show that central attitude elements have a stronger impact on behavioural decisions than peripheral attitude elements (e.g., your positive beliefs about Hillary Clinton were more central in your attitude network than your negative beliefs, so the chance that you voted for Hillary Clinton further increased). We thus provide insight into how structural properties of attitudes determine the extent to which attitudes have impact on behaviour.
 \par
\begin{figure*}[]
    \centering
\includegraphics[width=6.27in]{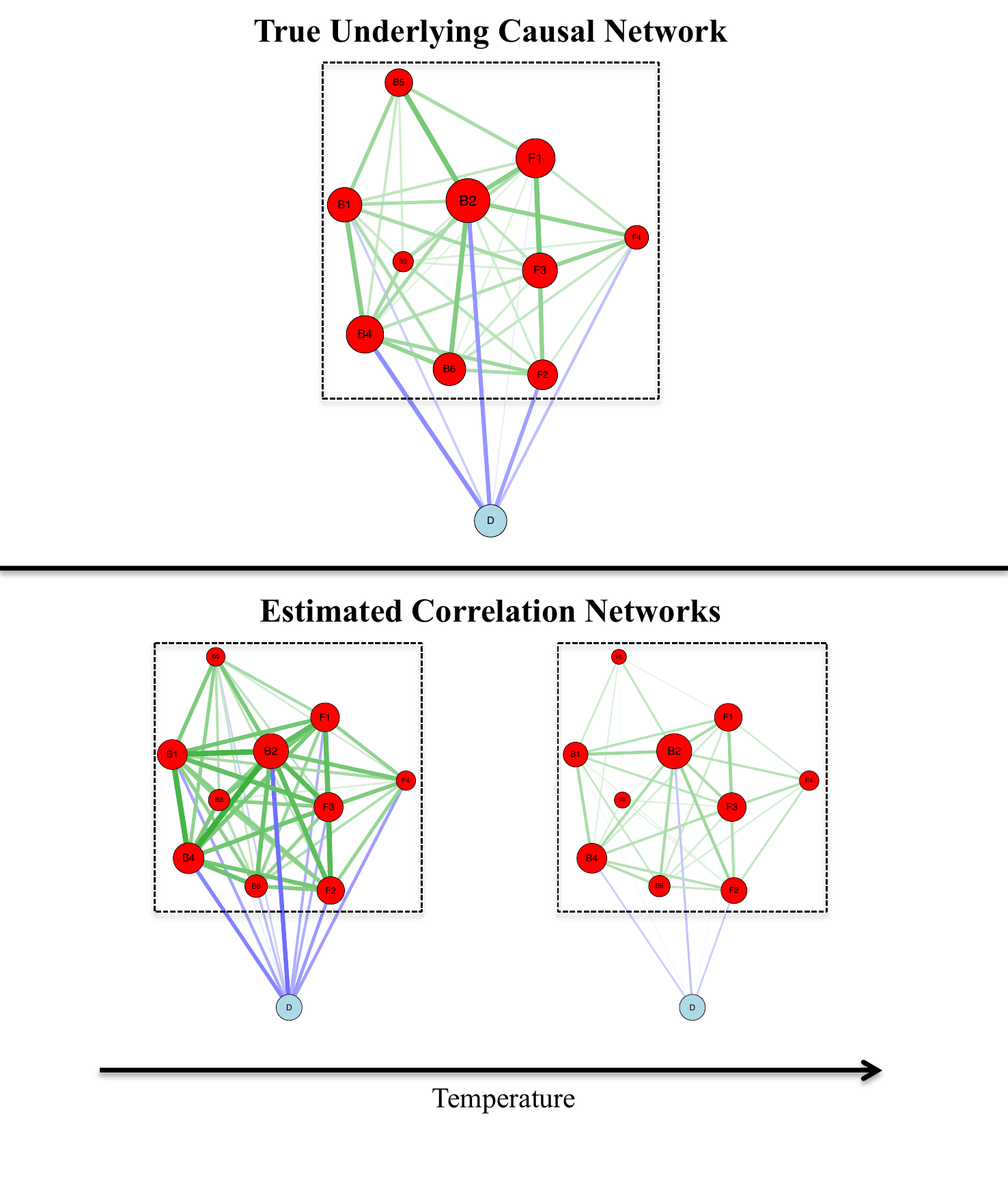}
\caption{\textbf{Illustrations of the Causal Attitude Network model and the hypotheses of the current study.} Networks represent a hypothetical attitude network toward a presidential candidate consisting of six beliefs (e.g., judging the candidate as honest, intelligent, caring; represented by nodes B1 to B6), four feelings (e.g., feeling hope, anger toward the candidate; represented by nodes F1 to F4), and the voting decision (represented by the node D). Red nodes within the dashed square represent the part of the network on which connectivity and centrality estimates are calculated. Edges represent positive bidirectional causal influences (correlations) in the causal network (correlation networks), with thicker edges representing higher influence (correlations). Note that in this network, we assume that positive (negative) states of all nodes indicate a positive (negative) evaluation (e.g., positive state of judging a candidate as honest (dishonest) would be to (not) endorse this judgment). Size of the red nodes corresponds to their closeness centrality (see Methods for details on the network descriptives). In the CAN model, temperature represents a formalized conceptualization of consistency pressure on attitude networks. The correlation networks illustrate that lower (higher) temperature implies higher (lower) correlations between the attitude elements.}
\label{fig:fig1}
\end{figure*}

In network theory, dynamical systems are modelled as a set of nodes, representing autonomous entities, and edges, representing interactions between the nodes \cite{Newman2010}. The set of nodes and edges jointly defines a network structure. Modelling complex systems in this way has probably become the most promising data-analytic tool to tackle complexity in many fields \cite{Barabasi2011}, such as physics \cite{Barabasi1999, Watts1998}, biology \cite{Barabasi2004}, and psychology \cite{Cramer2010, Cramer2012, vandeLemmput2014, vanBorkulo2015}. Recently, network analysis has also been introduced to the research on attitudes in the form of the Causal Attitude Network (CAN) model \cite{Dalege2016}. In this model, attitudes are conceptualized as networks, in which nodes represent attitude elements that are connected by direct causal interactions (see Figure \ref{fig:fig1}). The CAN model further assumes that the Ising model \cite{Ising1925}, which originated from statistical physics, represents an idealized model of attitude dynamics.\par

In the Ising model, the probability of configurations (i.e., the states of all nodes in the network), which represents the overall state of the attitude network, depends on the amount of \textit{energy} of a given configuration. The energy of a given configuration can be calculated using the Hamiltonian function:
\begin{equation} 
H(x)=-\sum_i\tau_ix_i-\sum_{<i,j>}\omega_ix_ix_j.
\end{equation}
Here, \textit{k} distinct attitude elements 1,....,\textit{i},\textit{j},....\textit{k} are represented as nodes that engage in pairwise interactions; the variables $x_i$ and $x_j$ represent the states of nodes \textit{i} and \textit{j} respectively. The model is designed to represent the probability of these states as a function of a number of parameters that encode the network structure. The parameter $\tau_i$ is the threshold of node \textit{i}, which determines the disposition of that node to be in a positive state (1; endorsing an attitude element) or negative state (-1, not endorsing an attitude element) regardless of the state of the other nodes in the network (statistically, this parameter functions as an intercept). The parameter $\omega_i$ represents the edge weight (i.e., the strength of interaction) between nodes \textit{i} and \textit{j}. As can be seen in this equation, the Hamiltonian energy decreases if nodes are in a state that is congruent with their threshold and when two nodes having positive (negative) edge weights assume the same (different) state. Assuming that attitude elements of the same (different) valence are generally positively (negatively) connected, attitude networks thus strive for a consistent representation of the attitude. The probability of a given configuration can be calculated using the Gibbs distribution \cite{Murphy2012}:
\begin{equation} 
\Pr(X=x)=\frac{exp(-\beta H(x)}{Z},
\end{equation}
in which $\beta$ represents the inverse temperature of the system, which can be seen as consistency pressures on attitude networks: reducing (increasing) the temperature of the system results in stronger (weaker) influence of the thresholds and weights, thereby scaling the entropy of the Ising network model \cite{Epskamp2016, Wainwright2008}. An Ising model with low (high) temperature results in a highly (weakly) connected correlation network (see Figure \ref{fig:fig1}). The denominator $Z$ represents the sum of the energies of all possible configurations, which acts as a normalising factor to ensure that the sum of the probabilities adds up to 1. \par
Conceptualising attitudes as Ising models allows for the derivation of several hypotheses and a crucial test of this conceptualisation is whether it can advance the understanding of the relation between attitudes and behavioural decisions. In the present paper we apply the CAN model and are the first to (a) formalize and (b) test hypotheses based on the CAN model regarding the impact of attitudes on behaviour.\par
The impact of attitudes on behaviour has been one of the central research themes in Social Psychology in recent decades \cite{Ajzen1991, Glasman2006, Kruglanski2015}. The bulk of the research on the relation between attitudes and behaviour has been done under the umbrella definition of attitude strength, which holds that one central feature of strong attitudes is that they have a strong impact on behaviour \cite{Krosnick1995}. Several lines of research have identified factors related to attitude strength. Among the most widely researched of these are attitude accessibility, attitude importance, and attitudinal ambivalence.  Studies have shown that accessible attitudes (i.e., attitudes that can be easily retrieved from memory) have more impact on behaviour \cite{Fazio1986, Glasman2006}. Similarly, higher levels of (subjective) attitude importance (i.e., attitudes, to which a person attaches subjective importance), are related to increased accessibility of attitudes \cite{Krosnick1989} and to higher levels of consistency between attitudes and behaviour \cite{Krosnick1988, Visser2003}. Ambivalent attitudes (i.e., attitudes that are based on both negative and positive associations) are less predictive of behaviour than univalent attitudes \cite{Armitage2000, vanHarreveld2015}. While these and other attitude strength attributes, such as certainty and extremity, are generally interrelated \cite{Krosnick1993, Visser2006}, a framework that unifies these different attributes has long been absent in the literature. Recently, however, based on the development of the CAN model, attitude strength was formally conceptualized as network connectivity \cite{Dalege2016}. The CAN model might thus provide the basis for a comprehensive and formalized framework of the relationship between attitudes and behaviour. Our current aim is to develop and test such a framework. To do so, we first formally derive hypotheses regarding the impact of attitudes on behaviour from the CAN model. Second, we test these hypotheses in the context of voting decisions in the US American presidential elections. \par
From the CAN model the hypothesis follows that highly connected attitude networks (i.e., attitude networks that are based on Ising models with low temperature) have a strong impact on behaviour. As can be seen in Figure \ref{fig:fig1}, low temperature results in strong connections both between non-behavioural attitude elements (i.e., beliefs and feelings) \textit{and} between non-behavioural attitude elements and behaviours (e.g., behavioural decisions) \cite{Dalege2016}. Attitude elements in highly connected networks are thus expected to have a strong impact on behavioural decisions. This leads to the hypothesis that the overall impact of attitudes depends on the connectivity of the attitude network. While the connectivity of attitude networks provides a novel formalisation of attitude strength, earlier approaches to understanding the structure of attitudes fit very well within this framework. For example, studies have shown that important attitudes are more coherent than unimportant attitudes \cite{Judd1981, Judd1989} and that strong attitudes have a more consistent structure between feelings and beliefs than weak attitudes \cite{Chaiken1995}. Also, Phillip E. Converse's  \citeyear{Converse1970} distinction between attitudes and nonattitudes based on stability of responses relates to our connectivity framework \cite{Dalege2016}\par
In addition to predicting the overall impact of an attitude from the connectivity of the attitude network, the CAN model predicts that the specific impact of attitude elements depends on their centrality (as defined by their closeness). Closeness refers to how strongly a given node is connected both directly and indirectly to all other nodes in the network \cite{Freeman1978, Opsahl2010}. In contrast to connectivity, which represents a measure of the whole network, centrality is a measure that applies to individual nodes within the network. Attitude elements high in closeness are good proxies of the overall state of the attitude network, as they hold more information about the rest of the network than peripheral attitude elements, rendering closeness the optimal measure of centrality for our current purposes. We therefore expect central attitude elements to have a stronger impact (directly or indirectly) on a behavioural decision no matter which attitude elements are direct causes of this decision. This can also be seen in Figure \ref{fig:fig1}, as there is a strong relation between a given node's centrality and it's correlation with the behavioural decision. It is important to note here that centrality of attitude elements does not refer to the classical definition of attitude centrality, but to the network analytical meaning of centrality. Specific impact of attitude elements has received somewhat less attention in the attitude literature than the global impact of attitudes, with studies either focusing on the primacy of feelings or beliefs in determining behaviour \cite{Galdi2008, Lavine1998, Millar1996} or on the subjective importance of attitude elements \cite{vanHarreveld2000, vanderPligt2000} and these different lines of research have been carried out much in isolation from each other and from the attitude strength research paradigm \cite<for an exception see>{vanHarreveld2000}. It is our view that an advantage of the approach we take in this article is that our framework holds promise in unifying these different approaches to understanding the relation between attitudes and behaviour.\par
In this paper, we first show that the hypotheses put forward here above directly follow from conceptualising attitudes as networks with a simulation study. We then test these hypotheses using data on attitudes toward candidates and voting in the American presidential elections from 1980-2012. In doing so, we test whether the CAN model provides a comprehensive framework on whether attitudes and which attitude elements drive behavioural decisions. Voting decisions are a perfect test of this postulate, because political attitudes often but not always drive voting decisions \cite{Fazio1986, Galdi2008, Lavine1998, Kraus1986, Markus1982}. 
\section*{Results} 
\subsection*{Simulation Study}
To show that the hypotheses presented above directly follow from conceptualising attitudes as networks, we simulated networks using three popular algorithms to generate networks: preferential attachment \cite{Barabasi1999, Albert2002}, small-world network model \cite{Watts1998}, and random Erdos-R\'enyi networks \cite{Erdos1959} (see also Supplementary Note 1 for analytical solutions). The networks consisted of 11 nodes (which corresponds to the number of nodes in the empirically estimated networks described below), with ten randomly chosen nodes representing attitude elements and one randomly chosen node representing the behavioural decision. Note that in such small networks, network properties other than density and magnitude of edge weights do not play a fundamental role in determining outcomes of the network.\par
The simulation of networks followed four steps: First, we created a 'base' network using one of the three algorithms. Second, we added edge weights to the base network, either drawn from a normal distribution, a Pareto power law distribution, or a uniform distribution. Third, to simulate responses of individuals holding attitudes with the network structure of the base network, we used the Ising network model \cite{Ising1925}. We created 20 different variations of the weighted base network in which the temperature of the Ising model was varied. Fourth, we simulated 1000 individuals based on the variations of the base network. As can be seen in Figure \ref{fig:fig1}, increasing (decreasing) the temperature results in decreasing (increasing) edge weights in the correlation networks.\par
We repeated this procedure 100 times for each combination of network generating algorithms and edge weights distributions. To investigate whether simulated attitude elements in highly connected networks (i.e., networks, for which the temperature parameter was low) collectively have a strong impact on the simulated decision, we estimated the global connectivity, defined by the Average Shortest Path Length \cite<ASPL,>{West1996} of the simulated attitude elements. We correlated the global connectivity with the average impact (which we operationalize as the biserial correlation between the sum score of the simulated attitude elements and the simulated decision) for each set of 20 networks. This resulted in strong negative correlations collapsed over all combinations of network-generating algorithms and edge weights distributions (Pearson correlations: mean \textit{r}=-0.91, s.d. \textit{r}=0.06) and we found strong negative correlations for all of these combinations (see Supplementary Table \ref{tab:tabS1}). To investigate whether central nodes (based on closeness) have a strong impact on a decision, we estimated the centrality of the simulated attitude elements and correlated the centrality estimates with the impact of the simulated attitude elements (which we operationalize as the tetrachoric correlation between a given simulated attitude element and the simulated decision). To exclude the possibility that results are driven by differences in average centrality and impact, we standardized both centrality and impact for each network. This resulted in strong positive correlations in the different sets of attitude networks collapsed over all combinations of network-generating algorithms and edge weights distributions (Pearson correlations: mean \textit{r}=0.59, s.d. \textit{r}=0.29) and we found strong positive correlations for all of these combinations (see Supplementary Table \ref{tab:tabS1}).

\begin{table*}[]
\small
\caption{\textbf{Included attitude elements.}}
\label{tab:tab1}
\begin{tabular}{lll}\hline
           \cline{1-3}
Attitude element & Included in data set & Substituted by  \\  \hline
"is honest"* & 1988--1996, 2008--2012 & "is dishonest"* (1980, 2000--2004), "is decent"* (1984) \\
"is intelligent"* & 1984--1992, 1996 (Clinton), 2000--2012 & "is weak"* (1980), "gets things done"* (1996 Dole) \\
"is knowledgeable"* & 1980--2012 & NA \\
"is moral"* & 1980--2012 & NA \\
"really cares about people like you"* & 1984--2012 & "is inspiring"* (1980) \\
"would provide strong leadership"* & 1980--2012 & NA \\
"angry"** &	 1980--2012 & NA \\
"afraid"** & 1980--2012 & NA \\
"hopeful"** & 1980--2012 & NA \\
"proud"** & 1980--2012 & NA \\
 \hline
\end{tabular}
\raggedright{\textit{Note}. *Of these 1,714 participants, 1,316 participants also participated during the election of 1992.}
\end{table*}

\subsection*{Test of Connectivity Hypothesis}
These simulations show clearly that the CAN model predicts a strong relation between network connectivity (node centrality) and the predictive utility of attitudes (attitude elements) in forecasting behaviour. This confirms that these intuitively derived hypotheses are indeed formal predictions that must follow if the CAN model is a valid model of attitudes.  To provide an empirical test of the hypotheses put forward here, we analysed data from the American National Election Studies (ANES) on the US presidential elections from 1980--2012 (total \textit{n}=16,988). In each ANES between ten and 24 attitude elements were assessed and we selected ten attitude elements for each election that were most similar to each other, see Table \ref{tab:tab1}. On these ten attitude elements, we estimated attitude networks for each of the two (three) main candidates for the elections in 1984--1992 and in 2000--2012 (in 1980 and 1996).  This gave us 20 attitude networks in total. Nodes in these networks represent attitude elements toward the given presidential candidate that were rated by the participants. Edges between the nodes represent zero-order polychoric correlations between the attitude elements. Note that because our networks are based on zero-order correlations, these networks only vary in magnitudes of edge weights and not in density, because correlation networks are always fully connected.\par
\begin{figure*}[]
    \centering
\includegraphics[width=6.27in]{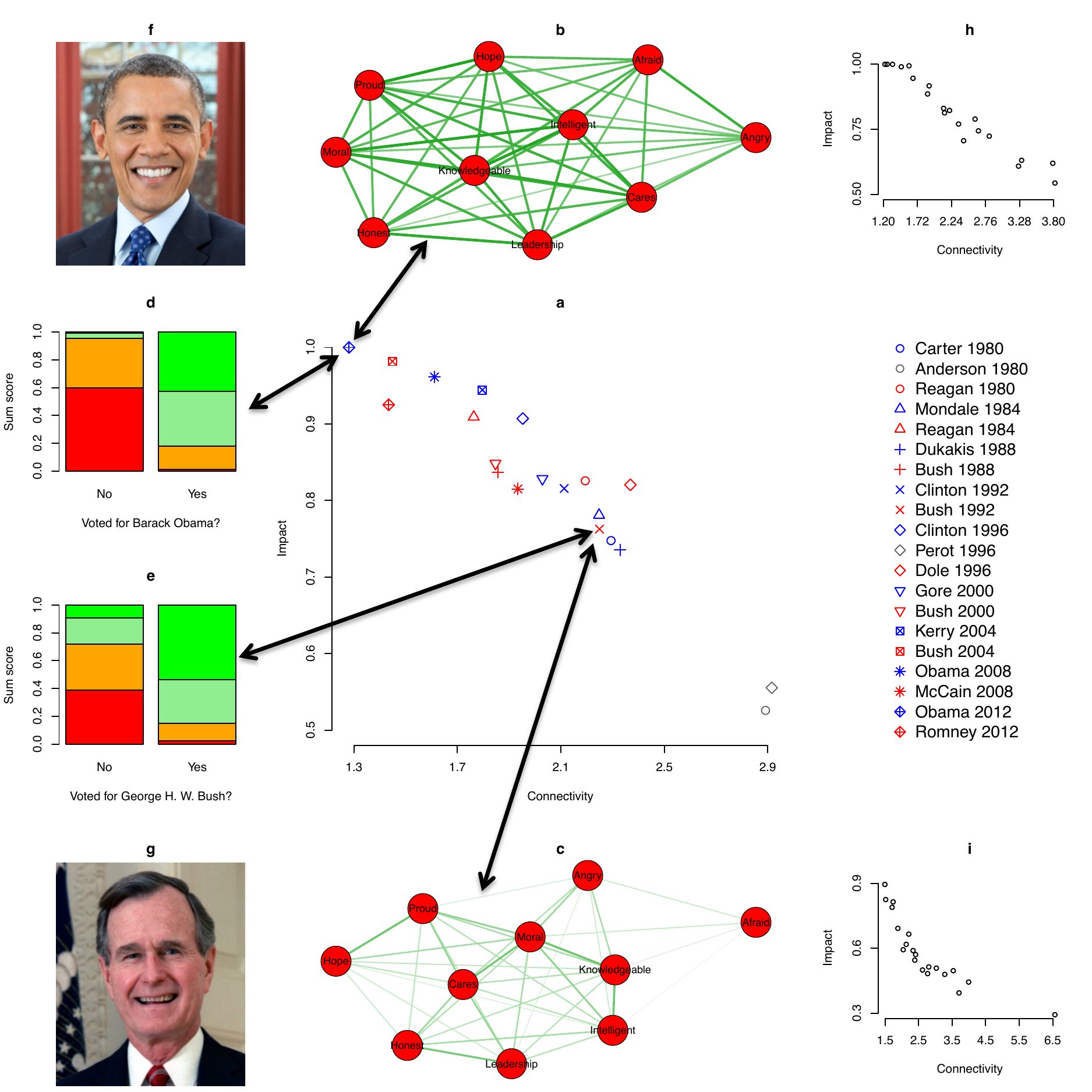}
\caption{\textbf{Highly connected attitude networks have a stronger impact on voting decisions than weakly connected attitude networks.} (\textbf{a}) Relation between connectivity and average impact of attitude elements. (\textbf{b})-(\textbf{e}) Two illustrations of the analytic strategy to assess connectivity and average impact. (b) [(c)] Attitude network toward Barack Obama [George H. W. Bush] in 2012 [1992]. Nodes represent attitude elements and edges represent correlations between attitude elements (the higher the correlation, the thicker the edge; correlations lower than .3 are not displayed). Closely connected attitude elements are placed near to each other \protect\cite{Fruchterman1991}. (\textbf{d}) [(\textbf{e})] Relation between the sum score of attitude elements toward Barack Obama [George H. W. Bush] and voting for Barack Obama [George H. W. Bush]. Colours of the bars represent the percentage of individuals who's sum scores fall into a given percentile (the more green, the higher the sum score; the more red, the lower the sum score). (\textbf{f}) Photo of Barack Obama by Pete Souza. Photo is under the CC0 / Public Domain Licence. Source: https://www.goodfreephotos.com/people/barack-obama-portrait-photo.jpg.php. (\textbf{g}) Photo of George H. W. Bush by unknown photographer. Photo is under the CC0 / Public Doman Licence. Source: https://www.goodfreephotos.com/people/george-bush-portrait-photo.jpg.php. (\textbf{h}) [(\textbf{i})] Relation between connectivity and impact for the simulated set of networks that was closest to the mean correlation plus [minus] one standard deviation.}
\label{fig:fig2}
\end{figure*}

First, we tested whether highly connected attitude networks have strong average impact. As in the simulation study, connectivity was based on the ASPL and average impact was operationalized as the biserial correlation between the sum score of attitude elements and the voting decision.  As can be seen in Figure \ref{fig:fig2}, we found a high negative correlation between connectivity and average impact (Pearson correlation: \textit{r}=-0.95, \textit{P}$<$0.001), supporting our hypothesis.\par
\subsection*{Test of Centrality Hypothesis}
\begin{figure*}[]
    \centering
\includegraphics[width=6.27in]{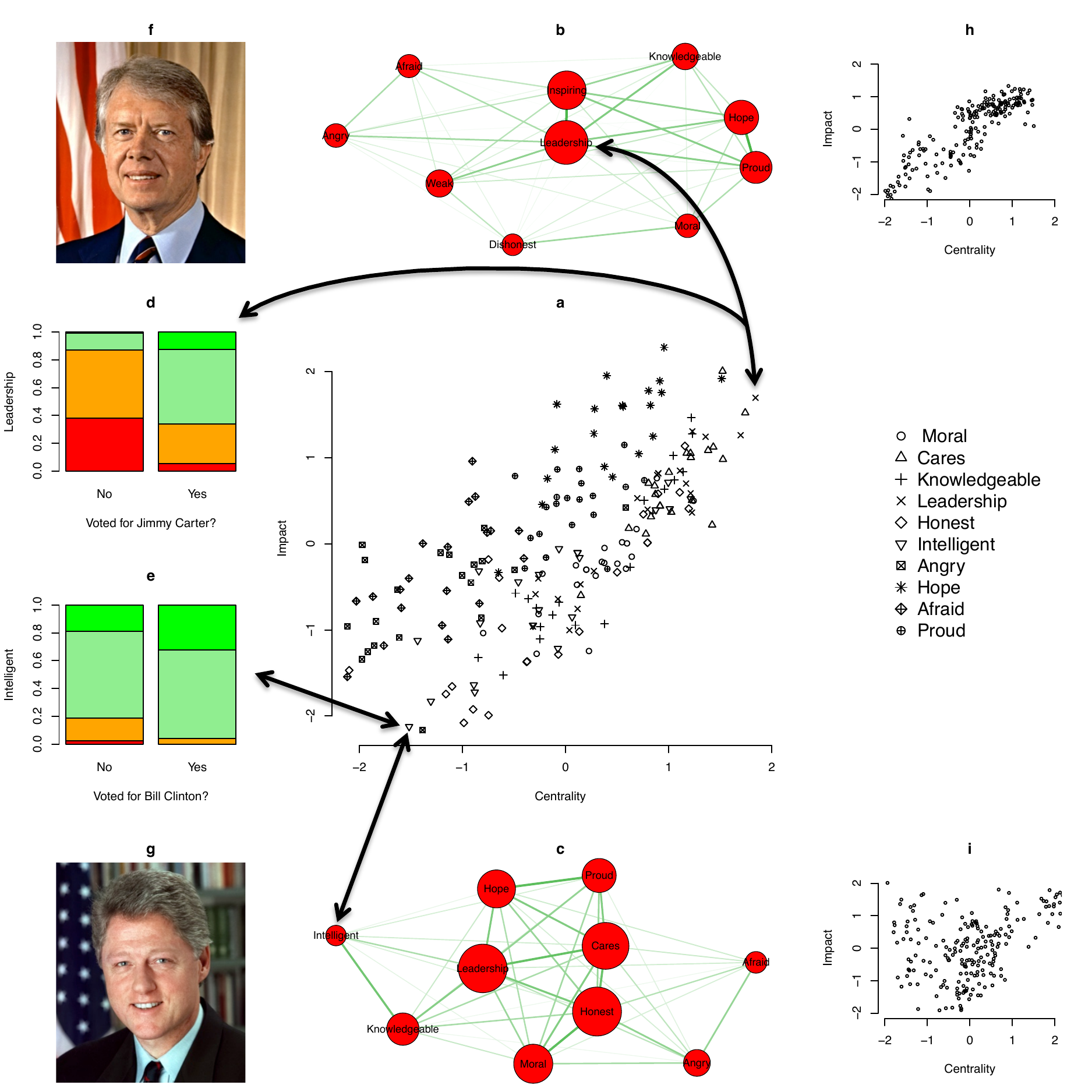}
\caption{\textbf{Central attitude elements have stronger impact on voting decisions than peripheral attitude elements.} (\textbf{a}) Relation between centrality and impact of attitude elements. (\textbf{b})-(\textbf{e}) Two illustrations of the analytic strategy to assess centrality and impact. (\textbf{b}) [(\textbf{c})] Attitude network toward Jimmy Carter [Bill Clinton] in 1980 [1992]. The networks have the same characteristics as the networks shown in Figure \ref{fig:fig2}, except that the size of the nodes corresponds to the nodes' relative centrality (the bigger the node, the higher its centrality). (\textbf{d}) [(\textbf{e}]) Relation between endorsing the belief that Jimmy Carter [Bill Clinton] would provide strong leadership [is intelligent] and voting for Jimmy Carter [Bill Clinton]. Colours of the bars represent the percentage of individuals who agree or do not agree with the judgment (the more green, the higher the agreement; the more red, the lower the agreement). See  Table \ref{tab:tab1} for more information on the attitude elements. (\textbf{f}) Photo of Jimmy Carter by unknown photographer. Photo is under the CC0 / Public Domain Licence. Source: https://www.goodfreephotos.com/people/jimmy-carter-portrait.jpg.php. (\textbf{g}) Photo of Bill Clinton by Bob McNeely. Photo is under the CC0 / Public Doman Licence. Source: https://www.goodfreephotos.com/people/bill-clinton-portrait-photo.jpg.php. (\textbf{h}) [(\textbf{i})] Relation between centrality and impact for the simulated set of networks that was closest to the mean correlation plus [minus] one standard deviation.}
\label{fig:fig3}
\end{figure*}

Second, we tested whether central attitude elements have a strong impact. This impact was operationalized as the polychoric correlation between a given attitude element and the voting decision.  We again standardized the centrality and impact estimates to exclude the possibility that results are driven by differences in mean centrality and mean impact. As can be seen in Figure \ref{fig:fig3}, we found a high positive correlation between standardized centrality and standardized impact (Pearson correlation: \textit{r}=0.70, \textit{P}$<$0.001), supporting our hypothesis.
\subsection*{Forecast Analysis}
To illustrate the practical relevance of our findings, we investigated whether centrality of attitude elements can be used to forecast the impact on voting decisions \textit{before} knowing the outcome of the election (e.g., whether our analyses can be used to forecast the impact of attitude elements on the next presidential election). For each election (e.g., election of 2012), we estimated the regression parameters between impact and centrality from all elections except the forecasted election (e.g., 1980-2008). We calculated the predicted impact in the forecasted election using the centrality indices of the forecasted election and the regression parameters. As can be seen in Figure \ref{fig:fig4}, the predicted impact was very close to the actual impact (deviation median=0.06, deviation interquartile range=0.03-0.09) and outperformed both using the mean of all attitude elements (Deviation median=0.12, deviation interquartile range=0.06-0.18, Wilcoxon-matched pairs test: \textit{V}=3346, \textit{P}$<$0.001, CLES=69.5\%) and using the means of the specific attitude elements (Deviation median=0.09, deviation interquartile range=0.04-0.17, Wilcoxon-matched pairs test: \textit{V}=5057, \textit{P}$<$0.001, CLES=65.2\%). Using centrality thus creates the possibility to forecast the (almost) exact impact of an attitude element on the voting decision.

\end{multicols}

\begin{multicols}{2}

\begin{center}
\captionsetup{type=figure}
\includegraphics[width=3.4in]{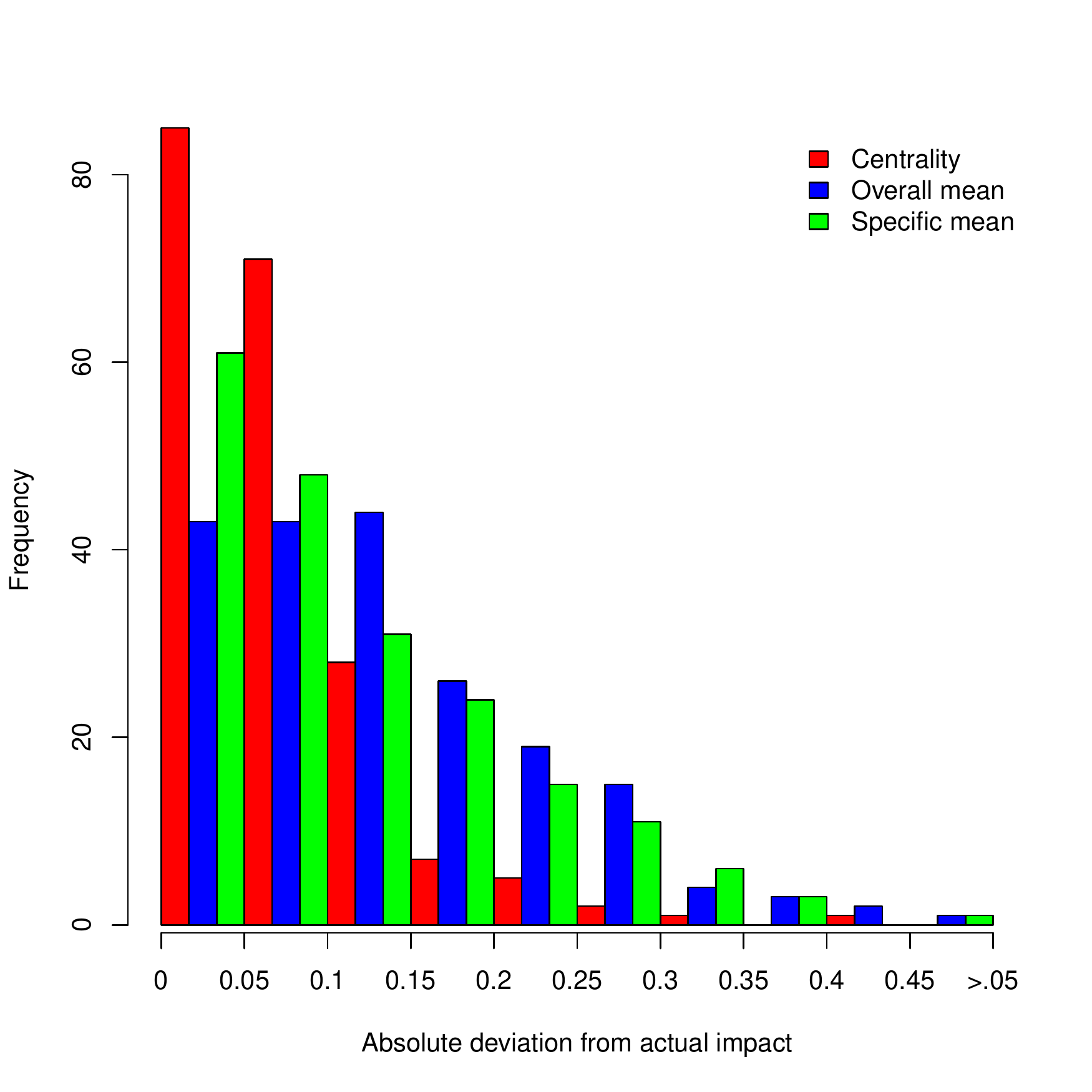}
  \caption{\textbf{Accuracy of forecasts based on centrality, overall mean, and specific mean.} The plot shows the results of forecasting the impact of each attitude element at each election.}
  \label{fig:fig4}

\end{center}

\section*{Discussion} 
Starting in the 1930s with Richard T. LaPiere's \citeyear{LaPiere1934} work, attitude-behaviour consistency has been one of the central research themes in Social Psychology \cite{Ajzen1991, Fazio1986, Fishbein1975, Glasman2006, Kraus1986, Wicker1969}. While early work focused on the question whether attitudes drive or do not drive behaviour \cite{Wicker1969}, more recent work has focused on when attitudes drive behaviour \cite{Glasman2006, Bagozzi1990, Ajzen1977, Fazio1986, Fazio1981}. This article provides a formalized and parsimonious answer to the question when attitudes drive behaviour: The impact of attitudes on behaviour depends on the connectivity of the attitude network, with central attitude elements having the highest impact on behaviour within a given attitude network.\par
The present research has shown that network structure of attitudes can inform election campaign strategies (and behavioural change programs in general) by predicting both the extent to which individuals base their decision on their attitude and the extent to which an attitude element influences voting decision (and other behaviour relevant to an attitude). Connectivity can help inform how effective candidate-centred campaigns would be. High connectivity indicates that voting decisions highly depend on candidate attitudes, while low connectivity indicates that other factors may play a more substantial role than candidate attitudes, such as party identification \cite{Bartels2000, Miller1991}, ideology \cite{Jacoby2010, Palfrey1987}, and public policy issues \cite{Abramowitz1995, Aldrich1989, Carmines1980, Fiorina1978, Nadeau2001, Rabinowitz1989}. Centrality can furthermore inform on the effectiveness of targeting specific attitude elements, as changing a central attitude element is probably more likely to affect the voting decision than changing a peripheral attitude element. \par
Future research might focus on how connectivity and centrality of attitude networks relate to other factors that influence voting decision. Among the most important factors influencing voting decisions are party identification \cite{Bartels2000, Miller1991} and specific policy issues \cite{Abramowitz1995, Aldrich1989, Carmines1980, Fiorina1978, Nadeau2001, Rabinowitz1989}. First, party identification might influence the connectivity of attitude networks, because it is likely that individuals, who identify with a political party, have a stronger drive for consistency in their attitudes toward presidential candidates. Party identification makes it also more likely that a given individual adopts a positive attitude toward the candidate of their party and it might also directly influence the voting decision. This makes party identification a possible confound of our results and we therefore also ran our analyses including only individuals, who do not identify with a political party. The results of this analysis mirrored the results of the results reported in this paper (see Supplementary Note 3 \& Supplementary Figure \ref{fig:figS3}). Second, policy issues might influence the centrality of attitude elements. If, for example, the current political climate is highly focused on foreign policies (e.g., the conflict in Syria), judging a candidate to be competent in respect to foreign policy making might take a central place in the attitude network. Generally, it is an important question for future research why some attitude elements are more central than others. Our analyses indicate that there are some attitude elements that are chronically central (see Figure \ref{fig:fig3}), with some variation that might be due to the specifics of the political climate during the different elections.\par
Another promising venture for future research would be to investigate how attitude networks develop during an election campaign. To do so, one could apply several intermediate assessments during the election campaign \cite{Erikson1999, Taleb2017, Wlezien2003, Dalege2017b}.The use of such intermediate assessment was shown to improve the prediction of election outcomes \cite{Taleb2017}. How might attitude networks change during an election? Based on the CAN model, we expect that (a) the connectivity of attitude networks heighten during an election campaign and (b) attitude networks probably grow due to the addition of newly formed attitude elements \cite{Dalege2016}. Also, predictions regarding the success of an election campaign to change a given person's attitude can be derived from the CAN model. Individuals holding attitudes that are based on highly connected networks already at the beginning of an election campaign are likely to not change their attitudes. Election campaigns might thus benefit from focusing on individuals holding attitudes that are based on weakly connected networks \cite{Dalege2017b}.\par    
In a broader sense, the CAN model advances our understanding of the relation between attitudes and behavioural decisions. Because the CAN model is a general model of attitudes, the results reported here likely generalize to other attitudes and behavioural decisions than those studied here as well. Using connectivity of attitude networks and centrality of attitude elements may for example provide more insight into issues such as which factors drive individuals to continue or stop smoking, buy a certain product, or behave aggressively toward a minority group. Furthermore, connectivity of attitude networks might unify the different approaches to explain variations in attitude-behaviour consistency, as it is likely that network connectivity is the glue that holds these factors together \cite{Dalege2016} and because our results indicate that network connectivity comprehensively explains variations in attitude-behaviour consistency. Several predictions above and beyond the findings reported here can also be derived from the network structure of attitudes. For example, network structure predicts when and which persuasion attempts will be successful \cite{Dalege2016}. Network theory thus holds great promise for advancing our understanding of the dynamical and structural properties of attitudes and their relation to a plethora of consequential human behaviours. 
\section*{Methods} 
\subsection*{Simulation of Networks} 
The simulation of networks followed four steps. First, an unweighted 'base' network consisting of 11 variables was created based on preferential attachment \cite{Barabasi1999, Albert2002}, the Small-World network model \cite{Watts1998}, or the Erdos-R\'enyi random graph model \cite{Erdos1959}  using the R package iGraph \cite{Csardi2006}. The preferential attachment algorithm starts with one node and then adds one node in each time step. The probability to which nodes the new node connects depends on the degree of the old node:
\begin{equation} 
\Pr(i)=\frac{k(i)^\alpha+1}{\sum_ik(j)^\alpha+1},
\end{equation}
where $k(i)$ is the degree of a given node. $\alpha$ was set to vary uniformly between 0.30 and 0.70. At each time step \textit{m} edges were added to the network. \textit{m} was set to vary uniformly between 4 and 6 \cite<resulting in relatively dense networks, as was shown to be the case for attitude networks,> {Dalege2016}. The Small-World network model starts with a ring lattice with nodes being connected to \textit{n} neighbours and then randomly rewires edges with a \textit{p} probability. \textit{n} was set to uniformly vary between 3 and 4 and \textit{p} was set to uniformly vary between 0.05 and 0.10. In the Erdos-R\'enyi graph, nodes are randomly connected by a given number of edges. Number of edges was set to uniformly vary between 30 and 45.\par
Second, edge weights were added to the base network. To have psychometrically realistic edge weights, we drew edge weights from either a normal distribution with \textit{M}=0.15 and \textit{SD}=.0075, a Pareto power law distribution with $\alpha$=3 and $\beta$=0.10, or a uniform distribution with range of 0.01--0.30.\par
Third, we created 20 variations of the weighted base network, in which the temperature of the Ising model was varied. The inverse temperature parameter $\beta$ was drawn from a normal distribution with \textit{M}=1 and \textit{SD}=0.2  (with higher numbers representing low entropy). To ensure that all nodes have roughly the same variance, we drew thresholds of nodes from a normal distribution with \textit{M}=0 and \textit{SD}=0.25.\par
Fourth, using the R-package IsingSampler \cite{Epskamp2016}, 1000 individuals for each of the variations of the base network were simulated based on the probability distribution implied by the Ising model. This procedure was repeated 900 times and each set of 20 variations of the different 900 base networks was analysed separately.\par
\subsection*{Participants}
The open-access data of the ANES involves large national random probability samples. Data were each collected in two interviews - one before and one after each presidential election from 1980 to 2012 - by the Center for Political Studies of the University of Michigan. In total, 21,365 participants participated in these nine studies (for \textit{Ns} per study see Supplementary Table \ref{tab:tabS2}), of which 16,667 participants stated that they voted for president. Non-voters were excluded from the analyses, because we assume that the decision whom to vote for is more likely to be part of the attitude network than the decision whether to vote or not. In Supplementary Note 2, however, we show that similar findings are obtained when non-voters are included in the analysis.\par  
\subsection*{Measures}
In each of the studies between six and 16 items tapping beliefs and between four and eight items tapping feelings toward the presidential candidates were assessed in the pre-election interviews. Feelings were assessed on two-point scales and beliefs were assessed on four-point scales (in a subsample of the ANES of 2008 and in the ANES of 2012, beliefs were assessed on a five-point scale). To have comparable attitude networks between the different elections, we always used six items tapping beliefs and four items tapping feelings (see  Table \ref{tab:tab1} for a list of included attitude elements). In the post-election interview, participants were asked which candidate they voted for. Depending on which presidential candidate the analysis focused, we scored the response as 1 when the participant stated that they voted for the given candidate and we scored the response as 0 when the participant did not vote for the given candidate.
\subsection*{Statistical Analyses}
We performed the same statistical analyses on the simulated and empirical data. 
\subsubsection*{Network Estimation}
Attitude networks were estimated using zero-order polychoric (tetrachoric) correlations between the (simulated) attitude elements as edge weights. We chose to use zero-order correlations as edge weights instead of estimating direct causal paths between the attitude elements because our simulations have shown that attitude networks based on zero-order correlations perform better than techniques that provide an estimate of the underlying causal network \cite{vanBorkulo2014}. 
\subsubsection*{Network Descriptives}
Both the ASPL and closeness are based on shortest path between lengths ($d$) between nodes. To calculate shortest path lengths, we used Dijkstra's algorithm  \cite{Dijkstra1959}, implemented in the R package qgraph \cite{Epskamp2012}:
\begin{equation}
d^w(i,j)=min(\frac{1}{w_{ih}}+\frac{1}{w_{hj}}).
\end{equation}
ASPL is then the average of the shortest path lengths between each pair of nodes in the network. Closeness ($c$) was calculated using the algorithm for weighted networks developed by \citeA{Opsahl2010}, using the R package qgraph:
\begin{equation}
c(i)=\Bigg[\sum^N_jd(i,j)\Bigg]^{-1}.
\end{equation}
\subsubsection*{Impact Estimates}
To estimate average impact of (simulated) attitude elements on (simulated) voting decisions, we calculated the biserial correlation between the sum score of (simulated) attitude elements and the (simulated) voting decision. We then calculated the Pearson correlation between connectivity and average impact for the 20 networks in the empirical study and for each set of 20 variations of the base networks in the simulation study. The clearly linear relation between connectivity and impact justified the use of Pearson correlation and significance testing. For the simulation study, we calculated the mean and standard deviation of the correlations obtained for each set of variations of the base network.\par
To estimate the impact of a given (simulated) attitude element on the (simulated) voting decision, we calculated the zero-order polychoric correlation between a given (simulated) attitude element and the (simulated) voting decision. We then calculated the Pearson correlation between standardized centrality and standardized impact of the attitude elements in the empirical study and for each set of 20 variations of the base networks in the simulation study. The clearly linear relation between centrality and impact justified the use of Pearson correlation and significance testing. For the simulation study, we calculated the mean and standard deviation of the correlations obtained for each set of variations of the base network.
\subsubsection*{Forecast Analysis}
For the forecast analysis, we first conducted nine regression analyses, in which impact was regressed on centrality. In each regression analysis, the forecasted election was omitted. From each of the regression equations, we first extracted the beta and intercept coefficients. Second, we multiplied the centrality indices of the forecasted election with the beta coefficient and added the intercept coefficient. Note that not in every election the same attitude elements were assessed. Of the ten used attitude elements, seven were assessed at each election. For the remaining three, we grouped the attitude elements together that were most similar to each other (see Table \ref{tab:tab1}). Third, we compared the resulting estimates with the actual impact of the attitude elements and calculated the absolute deviance scores. We then compared the performance of the centrality prediction to the overall mean prediction and the specific mean prediction. For both these predictions, we again calculated predictions nine times, omitting one of the elections each time. For the overall mean prediction, we calculated the mean of all attitude elements and for the specific mean prediction, we calculated the mean of each specific attitude element. We tested whether the centrality prediction performed better than the overall mean prediction and the specific mean prediction using Wilcoxon signed-rank tests.
\subsubsection*{Missing Values}
Missing values were deleted casewise (Supplementary Table \ref{tab:tabS2} shows the number of excluded participants per attitude network). Most missing values stemmed either from participants responding to an item that they did not know the answer or from non-participation during the post-election interview. Few missing values stemmed from interview errors. 
\subsubsection*{Alternative Analyses}
We also ran several alternative analyses that confirmed the robustness of our results: We ran alternative analyses on non-voters (see Supplementary Note 2 \& Supplementary Figure \ref{fig:figS2}), on independents (see Supplementary Note 3 \& Supplementary Figure \ref{fig:figS3}), on missing values (see Supplementary Note 4 \& Supplementary Figure \ref{fig:figS4}), on networks with different number of nodes (see Supplementary Note 5 \& Supplementary Figure \ref{fig:figS5}), and on latent variable models (see Supplementary Note 6 \& Supplementary Table \ref{tab:tabS3}).

\bibliography{bib} 

\begin{thebibliography}{}

\bibitem [\protect \citeauthoryear {%
Abramowitz%
}{%
Abramowitz%
}{%
{\protect \APACyear {1995}}%
}]{%
Abramowitz1995}
\APACinsertmetastar {%
Abramowitz1995}%
\begin{APACrefauthors}%
Abramowitz, A\BPBI I.%
\end{APACrefauthors}%
\unskip\
\newblock
\APACrefYearMonthDay{1995}{}{}.
\newblock
{\BBOQ}\APACrefatitle {It's abortion, stupid: Policy voting in the 1992
  presidential election} {It's abortion, stupid: Policy voting in the 1992
  presidential election}.{\BBCQ}
\newblock
\APACjournalVolNumPages{Journal of Politics}{57}{}{176--186}.
\PrintBackRefs{\CurrentBib}

\bibitem [\protect \citeauthoryear {%
Achen%
}{%
Achen%
}{%
{\protect \APACyear {1975}}%
}]{%
Achen1975}
\APACinsertmetastar {%
Achen1975}%
\begin{APACrefauthors}%
Achen, C.%
\end{APACrefauthors}%
\unskip\
\newblock
\APACrefYearMonthDay{1975}{}{}.
\newblock
{\BBOQ}\APACrefatitle {Mass political attitudes and the survey response} {Mass
  political attitudes and the survey response}.{\BBCQ}
\newblock
\APACjournalVolNumPages{American Political Science Review}{69}{}{1218--1231}.
\PrintBackRefs{\CurrentBib}

\bibitem [\protect \citeauthoryear {%
Ajzen%
}{%
Ajzen%
}{%
{\protect \APACyear {1991}}%
}]{%
Ajzen1991}
\APACinsertmetastar {%
Ajzen1991}%
\begin{APACrefauthors}%
Ajzen, I.%
\end{APACrefauthors}%
\unskip\
\newblock
\APACrefYearMonthDay{1991}{}{}.
\newblock
{\BBOQ}\APACrefatitle {The theory of planned behavior} {The theory of planned
  behavior}.{\BBCQ}
\newblock
\APACjournalVolNumPages{Organizational Behavior and Human Decision
  Processes}{50}{}{179--211}.
\PrintBackRefs{\CurrentBib}

\bibitem [\protect \citeauthoryear {%
Ajzen%
\ \BBA {} Fishbein%
}{%
Ajzen%
\ \BBA {} Fishbein%
}{%
{\protect \APACyear {1977}}%
}]{%
Ajzen1977}
\APACinsertmetastar {%
Ajzen1977}%
\begin{APACrefauthors}%
Ajzen, I.%
\BCBT {}\ \BBA {} Fishbein, M.%
\end{APACrefauthors}%
\unskip\
\newblock
\APACrefYearMonthDay{1977}{}{}.
\newblock
{\BBOQ}\APACrefatitle {Attitude-behavior relations: A theoretical analysis and
  review of empirical research} {Attitude-behavior relations: A theoretical
  analysis and review of empirical research}.{\BBCQ}
\newblock
\APACjournalVolNumPages{Psychological Bulletin}{84}{}{888--918}.
\PrintBackRefs{\CurrentBib}

\bibitem [\protect \citeauthoryear {%
Albert%
\ \BBA {} Barab\'asi%
}{%
Albert%
\ \BBA {} Barab\'asi%
}{%
{\protect \APACyear {2002}}%
}]{%
Albert2002}
\APACinsertmetastar {%
Albert2002}%
\begin{APACrefauthors}%
Albert, R.%
\BCBT {}\ \BBA {} Barab\'asi, A\BHBI L.%
\end{APACrefauthors}%
\unskip\
\newblock
\APACrefYearMonthDay{2002}{}{}.
\newblock
{\BBOQ}\APACrefatitle {Statistical mechanics of complex networks} {Statistical
  mechanics of complex networks}.{\BBCQ}
\newblock
\APACjournalVolNumPages{Review of Modern Physics}{74}{}{47--97}.
\PrintBackRefs{\CurrentBib}

\bibitem [\protect \citeauthoryear {%
Aldrich%
, Sullivan%
\BCBL {}\ \BBA {} Borgida%
}{%
Aldrich%
\ \protect \BOthers {.}}{%
{\protect \APACyear {1989}}%
}]{%
Aldrich1989}
\APACinsertmetastar {%
Aldrich1989}%
\begin{APACrefauthors}%
Aldrich, J\BPBI H.%
, Sullivan, J\BPBI L.%
\BCBL {}\ \BBA {} Borgida, E.%
\end{APACrefauthors}%
\unskip\
\newblock
\APACrefYearMonthDay{1989}{}{}.
\newblock
{\BBOQ}\APACrefatitle {Foreign affairs and issue voting: Do presidential
  candidates ''waltz before a blind audience?''} {Foreign affairs and issue
  voting: Do presidential candidates ''waltz before a blind
  audience?''}.{\BBCQ}
\newblock
\APACjournalVolNumPages{American Political Science Review}{83}{}{123--141}.
\PrintBackRefs{\CurrentBib}

\bibitem [\protect \citeauthoryear {%
Ansolabehere%
, Rodden%
\BCBL {}\ \BBA {} Snyder%
}{%
Ansolabehere%
\ \protect \BOthers {.}}{%
{\protect \APACyear {2008}}%
}]{%
Ansolabehere2008}
\APACinsertmetastar {%
Ansolabehere2008}%
\begin{APACrefauthors}%
Ansolabehere, S.%
, Rodden, J.%
\BCBL {}\ \BBA {} Snyder, J\BPBI M.%
\end{APACrefauthors}%
\unskip\
\newblock
\APACrefYearMonthDay{2008}{}{}.
\newblock
{\BBOQ}\APACrefatitle {The strength of issues: Using multiple measures to gauge
  preference stability, ideological constraint, and issue voting} {The strength
  of issues: Using multiple measures to gauge preference stability, ideological
  constraint, and issue voting}.{\BBCQ}
\newblock
\APACjournalVolNumPages{American Political Science Review}{102}{}{215--232}.
\PrintBackRefs{\CurrentBib}

\bibitem [\protect \citeauthoryear {%
Armitage%
\ \BBA {} Conner%
}{%
Armitage%
\ \BBA {} Conner%
}{%
{\protect \APACyear {2000}}%
}]{%
Armitage2000}
\APACinsertmetastar {%
Armitage2000}%
\begin{APACrefauthors}%
Armitage, C\BPBI J.%
\BCBT {}\ \BBA {} Conner, M.%
\end{APACrefauthors}%
\unskip\
\newblock
\APACrefYearMonthDay{2000}{}{}.
\newblock
{\BBOQ}\APACrefatitle {Attitudinal ambivalence: A test of three key hypotheses}
  {Attitudinal ambivalence: A test of three key hypotheses}.{\BBCQ}
\newblock
\APACjournalVolNumPages{Personality and Social Psychology
  Bulletin}{26}{}{1421--1432}.
\PrintBackRefs{\CurrentBib}

\bibitem [\protect \citeauthoryear {%
Bagozzi%
\ \BBA {} Baumgartner%
}{%
Bagozzi%
\ \BBA {} Baumgartner%
}{%
{\protect \APACyear {1986}}%
}]{%
Bagozzi1990}
\APACinsertmetastar {%
Bagozzi1990}%
\begin{APACrefauthors}%
Bagozzi, R.%
\BCBT {}\ \BBA {} Baumgartner, J.%
\end{APACrefauthors}%
\unskip\
\newblock
\APACrefYearMonthDay{1986}{}{}.
\newblock
{\BBOQ}\APACrefatitle {The level of effort required for behaviour as a
  moderator of the attitude-behaviour relation} {The level of effort required
  for behaviour as a moderator of the attitude-behaviour relation}.{\BBCQ}
\newblock
\APACjournalVolNumPages{European Journal of Social Psychology}{20}{}{45--59}.
\PrintBackRefs{\CurrentBib}

\bibitem [\protect \citeauthoryear {%
Barab\'asi%
}{%
Barab\'asi%
}{%
{\protect \APACyear {2011}}%
}]{%
Barabasi2011}
\APACinsertmetastar {%
Barabasi2011}%
\begin{APACrefauthors}%
Barab\'asi, A\BHBI L.%
\end{APACrefauthors}%
\unskip\
\newblock
\APACrefYearMonthDay{2011}{}{}.
\newblock
{\BBOQ}\APACrefatitle {The network takeover} {The network takeover}.{\BBCQ}
\newblock
\APACjournalVolNumPages{Nature Physics}{8}{}{14--16}.
\PrintBackRefs{\CurrentBib}

\bibitem [\protect \citeauthoryear {%
Barab\'asi%
\ \BBA {} Albert%
}{%
Barab\'asi%
\ \BBA {} Albert%
}{%
{\protect \APACyear {1999}}%
}]{%
Barabasi1999}
\APACinsertmetastar {%
Barabasi1999}%
\begin{APACrefauthors}%
Barab\'asi, A\BHBI L.%
\BCBT {}\ \BBA {} Albert, R.%
\end{APACrefauthors}%
\unskip\
\newblock
\APACrefYearMonthDay{1999}{}{}.
\newblock
{\BBOQ}\APACrefatitle {Emergence of scaling in random networks} {Emergence of
  scaling in random networks}.{\BBCQ}
\newblock
\APACjournalVolNumPages{Science}{286}{}{509--512}.
\PrintBackRefs{\CurrentBib}

\bibitem [\protect \citeauthoryear {%
Barab\'asi%
\ \BBA {} Zolt\'an%
}{%
Barab\'asi%
\ \BBA {} Zolt\'an%
}{%
{\protect \APACyear {2004}}%
}]{%
Barabasi2004}
\APACinsertmetastar {%
Barabasi2004}%
\begin{APACrefauthors}%
Barab\'asi, A\BHBI L.%
\BCBT {}\ \BBA {} Zolt\'an, N\BPBI O.%
\end{APACrefauthors}%
\unskip\
\newblock
\APACrefYearMonthDay{2004}{}{}.
\newblock
{\BBOQ}\APACrefatitle {Network biology: Understanding the cell's functional
  organization} {Network biology: Understanding the cell's functional
  organization}.{\BBCQ}
\newblock
\APACjournalVolNumPages{Nature Review Genetics}{5}{}{101--113}.
\PrintBackRefs{\CurrentBib}

\bibitem [\protect \citeauthoryear {%
Bartels%
}{%
Bartels%
}{%
{\protect \APACyear {2000}}%
}]{%
Bartels2000}
\APACinsertmetastar {%
Bartels2000}%
\begin{APACrefauthors}%
Bartels, L\BPBI M.%
\end{APACrefauthors}%
\unskip\
\newblock
\APACrefYearMonthDay{2000}{}{}.
\newblock
{\BBOQ}\APACrefatitle {Partisanship and voting behavior, 1952-1996}
  {Partisanship and voting behavior, 1952-1996}.{\BBCQ}
\newblock
\APACjournalVolNumPages{American Journal of Political Science}{44}{}{35--50}.
\PrintBackRefs{\CurrentBib}

\bibitem [\protect \citeauthoryear {%
Besag%
}{%
Besag%
}{%
{\protect \APACyear {1974}}%
}]{%
Besag1974}
\APACinsertmetastar {%
Besag1974}%
\begin{APACrefauthors}%
Besag, J.%
\end{APACrefauthors}%
\unskip\
\newblock
\APACrefYearMonthDay{1974}{}{}.
\newblock
{\BBOQ}\APACrefatitle {Spatial interaction and the statistical analysis of
  lattice systems} {Spatial interaction and the statistical analysis of lattice
  systems}.{\BBCQ}
\newblock
\APACjournalVolNumPages{Journal of the Royal Statistical Society. Series B
  (Methodological)}{36}{}{192--236}.
\PrintBackRefs{\CurrentBib}

\bibitem [\protect \citeauthoryear {%
Breckler%
}{%
Breckler%
}{%
{\protect \APACyear {1984}}%
}]{%
Breckler1984}
\APACinsertmetastar {%
Breckler1984}%
\begin{APACrefauthors}%
Breckler, S\BPBI J.%
\end{APACrefauthors}%
\unskip\
\newblock
\APACrefYearMonthDay{1984}{}{}.
\newblock
{\BBOQ}\APACrefatitle {Empirical validation of affect, behavior, and cognition
  as distinct components of attitude} {Empirical validation of affect,
  behavior, and cognition as distinct components of attitude}.{\BBCQ}
\newblock
\APACjournalVolNumPages{Journal of Personality and Social
  Psychology}{47}{}{1191--1205}.
\PrintBackRefs{\CurrentBib}

\bibitem [\protect \citeauthoryear {%
Brown%
}{%
Brown%
}{%
{\protect \APACyear {1986}}%
}]{%
Brown1986}
\APACinsertmetastar {%
Brown1986}%
\begin{APACrefauthors}%
Brown, L.%
\end{APACrefauthors}%
\unskip\
\newblock
\APACrefYear{1986}.
\newblock
\APACrefbtitle {Fundamentals of statistical exponential families} {Fundamentals
  of statistical exponential families}.
\newblock
\APACaddressPublisher{}{Hayward, CA:Institute of Mathematical Statistics}.
\PrintBackRefs{\CurrentBib}

\bibitem [\protect \citeauthoryear {%
Carmines%
\ \BBA {} Stimson%
}{%
Carmines%
\ \BBA {} Stimson%
}{%
{\protect \APACyear {1980}}%
}]{%
Carmines1980}
\APACinsertmetastar {%
Carmines1980}%
\begin{APACrefauthors}%
Carmines, E.%
\BCBT {}\ \BBA {} Stimson, J.%
\end{APACrefauthors}%
\unskip\
\newblock
\APACrefYearMonthDay{1980}{}{}.
\newblock
{\BBOQ}\APACrefatitle {The two faces of issue voting} {The two faces of issue
  voting}.{\BBCQ}
\newblock
\APACjournalVolNumPages{American Political Science Review}{74}{}{78--91}.
\PrintBackRefs{\CurrentBib}

\bibitem [\protect \citeauthoryear {%
Chaiken%
, Pomerantz%
\BCBL {}\ \BBA {} Giner-Sorolla%
}{%
Chaiken%
\ \protect \BOthers {.}}{%
{\protect \APACyear {1995}}%
}]{%
Chaiken1995}
\APACinsertmetastar {%
Chaiken1995}%
\begin{APACrefauthors}%
Chaiken, S.%
, Pomerantz, E\BPBI M.%
\BCBL {}\ \BBA {} Giner-Sorolla, R.%
\end{APACrefauthors}%
\unskip\
\newblock
\APACrefYearMonthDay{1995}{}{}.
\newblock
{\BBOQ}\APACrefatitle {Structural consistency and attitude strength}
  {Structural consistency and attitude strength}.{\BBCQ}
\newblock
\BIn{} R\BPBI E.~Petty\ \BBA {} J\BPBI A.~Krosnick\ (\BEDS), \APACrefbtitle
  {Attitude strength: Antecedents and consequences} {Attitude strength:
  Antecedents and consequences}\ (\BPGS\ 387--412).
\newblock
\APACaddressPublisher{}{Hillsdale, NJ: Lawrence Erlbaum}.
\PrintBackRefs{\CurrentBib}

\bibitem [\protect \citeauthoryear {%
Converse%
}{%
Converse%
}{%
{\protect \APACyear {1970}}%
}]{%
Converse1970}
\APACinsertmetastar {%
Converse1970}%
\begin{APACrefauthors}%
Converse, P\BPBI E.%
\end{APACrefauthors}%
\unskip\
\newblock
\APACrefYearMonthDay{1970}{}{}.
\newblock
{\BBOQ}\APACrefatitle {Attitudes and non-attitudes: Continuation of a dialogue}
  {Attitudes and non-attitudes: Continuation of a dialogue}.{\BBCQ}
\newblock
\BIn{} E\BPBI R.~Tufte\ (\BED), \APACrefbtitle {The quantative analysis of
  social problems} {The quantative analysis of social problems}\ (\BPGS\
  168--189).
\newblock
\APACaddressPublisher{}{Reading, MA: Addison-Wesley}.
\PrintBackRefs{\CurrentBib}

\bibitem [\protect \citeauthoryear {%
Cramer%
\ \protect \BOthers {.}}{%
Cramer%
\ \protect \BOthers {.}}{%
{\protect \APACyear {2012}}%
}]{%
Cramer2012}
\APACinsertmetastar {%
Cramer2012}%
\begin{APACrefauthors}%
Cramer, A\BPBI O\BPBI J.%
, van~der Sluis, S.%
, Noordhof, A.%
, Wichers, M.%
, Geschwind, N.%
, Aggen, S\BPBI H.%
\BDBL {}Borsboom, D.%
\end{APACrefauthors}%
\unskip\
\newblock
\APACrefYearMonthDay{2012}{}{}.
\newblock
{\BBOQ}\APACrefatitle {Dimensions of normal personality as networks in search
  of equilibrium: You can't like parties if you don't like people} {Dimensions
  of normal personality as networks in search of equilibrium: You can't like
  parties if you don't like people}.{\BBCQ}
\newblock
\APACjournalVolNumPages{European Journal of Personality}{26}{}{414--431}.
\PrintBackRefs{\CurrentBib}

\bibitem [\protect \citeauthoryear {%
Cramer%
, Waldorp%
, van~der Maas%
\BCBL {}\ \BBA {} Borsboom%
}{%
Cramer%
\ \protect \BOthers {.}}{%
{\protect \APACyear {2010}}%
}]{%
Cramer2010}
\APACinsertmetastar {%
Cramer2010}%
\begin{APACrefauthors}%
Cramer, A\BPBI O\BPBI J.%
, Waldorp, L\BPBI J.%
, van~der Maas, H\BPBI L\BPBI J.%
\BCBL {}\ \BBA {} Borsboom, D.%
\end{APACrefauthors}%
\unskip\
\newblock
\APACrefYearMonthDay{2010}{}{}.
\newblock
{\BBOQ}\APACrefatitle {Comorbidity: A network perspective} {Comorbidity: A
  network perspective}.{\BBCQ}
\newblock
\APACjournalVolNumPages{Behavioral and Brain Sciences}{33}{}{137--193}.
\PrintBackRefs{\CurrentBib}

\bibitem [\protect \citeauthoryear {%
Cs\'ardi%
\ \BBA {} Nepusz%
}{%
Cs\'ardi%
\ \BBA {} Nepusz%
}{%
{\protect \APACyear {2006}}%
}]{%
Csardi2006}
\APACinsertmetastar {%
Csardi2006}%
\begin{APACrefauthors}%
Cs\'ardi, G.%
\BCBT {}\ \BBA {} Nepusz, T.%
\end{APACrefauthors}%
\unskip\
\newblock
\APACrefYearMonthDay{2006}{}{}.
\newblock
{\BBOQ}\APACrefatitle {The igraph software package for complex network
  research} {The igraph software package for complex network research}.{\BBCQ}
\newblock
\APACjournalVolNumPages{InterJournal Complex Systems}{1695}{}{}.
\PrintBackRefs{\CurrentBib}

\bibitem [\protect \citeauthoryear {%
Dalege%
\ \protect \BOthers {.}}{%
Dalege%
\ \protect \BOthers {.}}{%
{\protect \APACyear {2016}}%
}]{%
Dalege2016}
\APACinsertmetastar {%
Dalege2016}%
\begin{APACrefauthors}%
Dalege, J.%
, Borsboom, D.%
, van Harreveld, F.%
, van~den Berg, H.%
, Conner, M.%
\BCBL {}\ \BBA {} van~der Maas, H\BPBI L\BPBI J.%
\end{APACrefauthors}%
\unskip\
\newblock
\APACrefYearMonthDay{2016}{}{}.
\newblock
{\BBOQ}\APACrefatitle {{Toward a formalized account of attitudes: The Causal
  Attitude Network (CAN) model}} {{Toward a formalized account of attitudes:
  The Causal Attitude Network (CAN) model}}.{\BBCQ}
\newblock
\APACjournalVolNumPages{Psychological Review}{123}{}{2--22}.
\PrintBackRefs{\CurrentBib}

\bibitem [\protect \citeauthoryear {%
Dalege%
, Borsboom%
, van Harreveld%
\BCBL {}\ \BBA {} van~der Maas%
}{%
Dalege%
\ \protect \BOthers {.}}{%
{\protect \APACyear {2017}}%
}]{%
Dalege2017b}
\APACinsertmetastar {%
Dalege2017b}%
\begin{APACrefauthors}%
Dalege, J.%
, Borsboom, D.%
, van Harreveld, F.%
\BCBL {}\ \BBA {} van~der Maas, H\BPBI L\BPBI J.%
\end{APACrefauthors}%
\unskip\
\newblock
\APACrefYearMonthDay{2017}{}{}.
\newblock
\APACrefbtitle {A network perspective on political attitudes: Testing the
  connectivity hypothesis.} {A network perspective on political attitudes:
  Testing the connectivity hypothesis.}
\newblock
\APACrefnote{https://arxiv.org/abs/1705.00193}
\PrintBackRefs{\CurrentBib}

\bibitem [\protect \citeauthoryear {%
Dijkstra%
}{%
Dijkstra%
}{%
{\protect \APACyear {1959}}%
}]{%
Dijkstra1959}
\APACinsertmetastar {%
Dijkstra1959}%
\begin{APACrefauthors}%
Dijkstra, E\BPBI W.%
\end{APACrefauthors}%
\unskip\
\newblock
\APACrefYearMonthDay{1959}{}{}.
\newblock
{\BBOQ}\APACrefatitle {A note on two problems in connexion with graphs} {A note
  on two problems in connexion with graphs}.{\BBCQ}
\newblock
\APACjournalVolNumPages{Numerische Mathematik}{1}{}{269--271}.
\PrintBackRefs{\CurrentBib}

\bibitem [\protect \citeauthoryear {%
Eagly%
\ \BBA {} Chaiken%
}{%
Eagly%
\ \BBA {} Chaiken%
}{%
{\protect \APACyear {1993}}%
}]{%
Eagly1993}
\APACinsertmetastar {%
Eagly1993}%
\begin{APACrefauthors}%
Eagly, A\BPBI H.%
\BCBT {}\ \BBA {} Chaiken, S.%
\end{APACrefauthors}%
\unskip\
\newblock
\APACrefYear{1993}.
\newblock
\APACrefbtitle {The psychology of attitudes} {The psychology of attitudes}.
\newblock
\APACaddressPublisher{}{Orlando, FL: Harcourt Brace Jovanovich}.
\PrintBackRefs{\CurrentBib}

\bibitem [\protect \citeauthoryear {%
Epskamp%
, Cramer%
, Waldorp%
, Schmittmann%
\BCBL {}\ \BBA {} Borsboom%
}{%
Epskamp%
\ \protect \BOthers {.}}{%
{\protect \APACyear {2012}}%
}]{%
Epskamp2012}
\APACinsertmetastar {%
Epskamp2012}%
\begin{APACrefauthors}%
Epskamp, S.%
, Cramer, A\BPBI O\BPBI J.%
, Waldorp, L\BPBI J.%
, Schmittmann, V\BPBI D.%
\BCBL {}\ \BBA {} Borsboom, D.%
\end{APACrefauthors}%
\unskip\
\newblock
\APACrefYearMonthDay{2012}{}{}.
\newblock
{\BBOQ}\APACrefatitle {qgraph: Network visualizations of relationships in
  psychometric data} {qgraph: Network visualizations of relationships in
  psychometric data}.{\BBCQ}
\newblock
\APACjournalVolNumPages{Journal of Statistical Software}{48}{}{1--18}.
\PrintBackRefs{\CurrentBib}

\bibitem [\protect \citeauthoryear {%
Epskamp%
, Maris%
, Waldorp%
\BCBL {}\ \BBA {} Borsboom%
}{%
Epskamp%
\ \protect \BOthers {.}}{%
{\protect \APACyear {2016}}%
}]{%
Epskamp2016}
\APACinsertmetastar {%
Epskamp2016}%
\begin{APACrefauthors}%
Epskamp, S.%
, Maris, G.%
, Waldorp, L.%
\BCBL {}\ \BBA {} Borsboom, D.%
\end{APACrefauthors}%
\unskip\
\newblock
\APACrefYearMonthDay{2016}{}{}.
\newblock
\APACrefbtitle {Network psychometrics.} {Network psychometrics.}
\newblock
\APACrefnote{Preprint at https://arxiv.org/abs/1609.02818}
\PrintBackRefs{\CurrentBib}

\bibitem [\protect \citeauthoryear {%
Erdos%
\ \BBA {} R\'enyi%
}{%
Erdos%
\ \BBA {} R\'enyi%
}{%
{\protect \APACyear {1959}}%
}]{%
Erdos1959}
\APACinsertmetastar {%
Erdos1959}%
\begin{APACrefauthors}%
Erdos, P.%
\BCBT {}\ \BBA {} R\'enyi, A.%
\end{APACrefauthors}%
\unskip\
\newblock
\APACrefYearMonthDay{1959}{}{}.
\newblock
{\BBOQ}\APACrefatitle {On random graphs} {On random graphs}.{\BBCQ}
\newblock
\APACjournalVolNumPages{Publicationes Mathematicae}{6}{}{290--297}.
\PrintBackRefs{\CurrentBib}

\bibitem [\protect \citeauthoryear {%
Erikson%
\ \BBA {} Wlezien%
}{%
Erikson%
\ \BBA {} Wlezien%
}{%
{\protect \APACyear {1999}}%
}]{%
Erikson1999}
\APACinsertmetastar {%
Erikson1999}%
\begin{APACrefauthors}%
Erikson, R\BPBI S.%
\BCBT {}\ \BBA {} Wlezien, C.%
\end{APACrefauthors}%
\unskip\
\newblock
\APACrefYearMonthDay{1999}{}{}.
\newblock
{\BBOQ}\APACrefatitle {Presidential polls as a time series: The case of 1996}
  {Presidential polls as a time series: The case of 1996}.{\BBCQ}
\newblock
\APACjournalVolNumPages{Public Opinion Quarterly}{63}{}{163--177}.
\PrintBackRefs{\CurrentBib}

\bibitem [\protect \citeauthoryear {%
Fazio%
}{%
Fazio%
}{%
{\protect \APACyear {1995}}%
}]{%
Fazio1995}
\APACinsertmetastar {%
Fazio1995}%
\begin{APACrefauthors}%
Fazio, R\BPBI H.%
\end{APACrefauthors}%
\unskip\
\newblock
\APACrefYearMonthDay{1995}{}{}.
\newblock
{\BBOQ}\APACrefatitle {Attitudes as object-evaluation associations:
  Determinants, consequences, and correlates of attitude accessibility}
  {Attitudes as object-evaluation associations: Determinants, consequences, and
  correlates of attitude accessibility}.{\BBCQ}
\newblock
\BIn{} R\BPBI E.~Petty\ \BBA {} J\BPBI A.~Krosnick\ (\BEDS), \APACrefbtitle
  {Attitude strength: Antecedents and consequences} {Attitude strength:
  Antecedents and consequences}\ (\BPGS\ 247--282).
\newblock
\APACaddressPublisher{}{Hillsdale, NJ: Lawrence Erlbaum}.
\PrintBackRefs{\CurrentBib}

\bibitem [\protect \citeauthoryear {%
Fazio%
\ \BBA {} Williams%
}{%
Fazio%
\ \BBA {} Williams%
}{%
{\protect \APACyear {1986}}%
}]{%
Fazio1986}
\APACinsertmetastar {%
Fazio1986}%
\begin{APACrefauthors}%
Fazio, R\BPBI H.%
\BCBT {}\ \BBA {} Williams, C\BPBI J.%
\end{APACrefauthors}%
\unskip\
\newblock
\APACrefYearMonthDay{1986}{}{}.
\newblock
{\BBOQ}\APACrefatitle {Attitude accessibility as a moderator of the
  attitude-perception and attitude-behavior relations: An investigation of the
  1984 presidential election} {Attitude accessibility as a moderator of the
  attitude-perception and attitude-behavior relations: An investigation of the
  1984 presidential election}.{\BBCQ}
\newblock
\APACjournalVolNumPages{Journal of Personality and Social
  Psychology}{51}{}{505--514}.
\PrintBackRefs{\CurrentBib}

\bibitem [\protect \citeauthoryear {%
Fazio%
\ \BBA {} Zanna%
}{%
Fazio%
\ \BBA {} Zanna%
}{%
{\protect \APACyear {1981}}%
}]{%
Fazio1981}
\APACinsertmetastar {%
Fazio1981}%
\begin{APACrefauthors}%
Fazio, R\BPBI H.%
\BCBT {}\ \BBA {} Zanna, M.%
\end{APACrefauthors}%
\unskip\
\newblock
\APACrefYearMonthDay{1981}{}{}.
\newblock
{\BBOQ}\APACrefatitle {Direct experience and attitude-behavior consistency}
  {Direct experience and attitude-behavior consistency}.{\BBCQ}
\newblock
\APACjournalVolNumPages{Advances in Experimental Social
  Psychology}{14}{}{161--202}.
\PrintBackRefs{\CurrentBib}

\bibitem [\protect \citeauthoryear {%
Fishbein%
\ \BBA {} Ajzen%
}{%
Fishbein%
\ \BBA {} Ajzen%
}{%
{\protect \APACyear {1975}}%
}]{%
Fishbein1975}
\APACinsertmetastar {%
Fishbein1975}%
\begin{APACrefauthors}%
Fishbein, M.%
\BCBT {}\ \BBA {} Ajzen, I.%
\end{APACrefauthors}%
\unskip\
\newblock
\APACrefYear{1975}.
\newblock
\APACrefbtitle {Belief, attitude, intention and behavior: An introduction to
  theory and research} {Belief, attitude, intention and behavior: An
  introduction to theory and research}.
\newblock
\APACaddressPublisher{}{Reading, MA: Addison-Wesley}.
\PrintBackRefs{\CurrentBib}

\bibitem [\protect \citeauthoryear {%
Freeman%
}{%
Freeman%
}{%
{\protect \APACyear {1978}}%
}]{%
Freeman1978}
\APACinsertmetastar {%
Freeman1978}%
\begin{APACrefauthors}%
Freeman, L\BPBI C.%
\end{APACrefauthors}%
\unskip\
\newblock
\APACrefYearMonthDay{1978}{}{}.
\newblock
{\BBOQ}\APACrefatitle {Centrality in social networks: Conceptual clarification}
  {Centrality in social networks: Conceptual clarification}.{\BBCQ}
\newblock
\APACjournalVolNumPages{Social Networks}{1}{}{215--239}.
\PrintBackRefs{\CurrentBib}

\bibitem [\protect \citeauthoryear {%
Fruchterman%
\ \BBA {} Reingold%
}{%
Fruchterman%
\ \BBA {} Reingold%
}{%
{\protect \APACyear {1991}}%
}]{%
Fruchterman1991}
\APACinsertmetastar {%
Fruchterman1991}%
\begin{APACrefauthors}%
Fruchterman, T.%
\BCBT {}\ \BBA {} Reingold, E.%
\end{APACrefauthors}%
\unskip\
\newblock
\APACrefYearMonthDay{1991}{}{}.
\newblock
{\BBOQ}\APACrefatitle {Graph drawing by force directed placement} {Graph
  drawing by force directed placement}.{\BBCQ}
\newblock
\APACjournalVolNumPages{Software, Practice \& Experience}{21}{}{1129--1164}.
\PrintBackRefs{\CurrentBib}

\bibitem [\protect \citeauthoryear {%
Galdi%
, Arcuri%
\BCBL {}\ \BBA {} Gawronski%
}{%
Galdi%
\ \protect \BOthers {.}}{%
{\protect \APACyear {2008}}%
}]{%
Galdi2008}
\APACinsertmetastar {%
Galdi2008}%
\begin{APACrefauthors}%
Galdi, S.%
, Arcuri, L.%
\BCBL {}\ \BBA {} Gawronski, B.%
\end{APACrefauthors}%
\unskip\
\newblock
\APACrefYearMonthDay{2008}{}{}.
\newblock
{\BBOQ}\APACrefatitle {Automatic Mental Associations Predict Future Choices of
  Undecided Decision-Makers} {Automatic mental associations predict future
  choices of undecided decision-makers}.{\BBCQ}
\newblock
\APACjournalVolNumPages{Science}{321}{}{1100--1102}.
\PrintBackRefs{\CurrentBib}

\bibitem [\protect \citeauthoryear {%
Glasman%
\ \BBA {} Albarrac\'in%
}{%
Glasman%
\ \BBA {} Albarrac\'in%
}{%
{\protect \APACyear {2006}}%
}]{%
Glasman2006}
\APACinsertmetastar {%
Glasman2006}%
\begin{APACrefauthors}%
Glasman, L\BPBI R.%
\BCBT {}\ \BBA {} Albarrac\'in, D.%
\end{APACrefauthors}%
\unskip\
\newblock
\APACrefYearMonthDay{2006}{}{}.
\newblock
{\BBOQ}\APACrefatitle {Forming attitudes that predict future behavior: A
  meta-analysis of the attitude-behavior relation} {Forming attitudes that
  predict future behavior: A meta-analysis of the attitude-behavior
  relation}.{\BBCQ}
\newblock
\APACjournalVolNumPages{Psychological Bulletin}{132}{}{778--822}.
\PrintBackRefs{\CurrentBib}

\bibitem [\protect \citeauthoryear {%
Hastie%
, Tibshirani%
\BCBL {}\ \BBA {} Friedman%
}{%
Hastie%
\ \protect \BOthers {.}}{%
{\protect \APACyear {2001}}%
}]{%
Hastie2001}
\APACinsertmetastar {%
Hastie2001}%
\begin{APACrefauthors}%
Hastie, T.%
, Tibshirani, R.%
\BCBL {}\ \BBA {} Friedman, J.%
\end{APACrefauthors}%
\unskip\
\newblock
\APACrefYear{2001}.
\newblock
\APACrefbtitle {The elements of statistical learning} {The elements of
  statistical learning}.
\newblock
\APACaddressPublisher{}{New York, NY: Springer}.
\PrintBackRefs{\CurrentBib}

\bibitem [\protect \citeauthoryear {%
Hastie%
, Tibshirani%
\BCBL {}\ \BBA {} Wainwright%
}{%
Hastie%
\ \protect \BOthers {.}}{%
{\protect \APACyear {2015}}%
}]{%
Hastie2015}
\APACinsertmetastar {%
Hastie2015}%
\begin{APACrefauthors}%
Hastie, T.%
, Tibshirani, R.%
\BCBL {}\ \BBA {} Wainwright, M.%
\end{APACrefauthors}%
\unskip\
\newblock
\APACrefYear{2015}.
\newblock
\APACrefbtitle {Statistical Learning with Sparsity: The Lasso and
  Generalizations} {Statistical learning with sparsity: The lasso and
  generalizations}.
\newblock
\APACaddressPublisher{}{New York, NY: CRC Press}.
\PrintBackRefs{\CurrentBib}

\bibitem [\protect \citeauthoryear {%
Ising%
}{%
Ising%
}{%
{\protect \APACyear {1925}}%
}]{%
Ising1925}
\APACinsertmetastar {%
Ising1925}%
\begin{APACrefauthors}%
Ising, E.%
\end{APACrefauthors}%
\unskip\
\newblock
\APACrefYearMonthDay{1925}{}{}.
\newblock
{\BBOQ}\APACrefatitle {{Beitrag zur Theorie des Ferromagnetismus}} {{Beitrag
  zur Theorie des Ferromagnetismus}}.{\BBCQ}
\newblock
\APACjournalVolNumPages{Zeitschrift f\"ur Physik}{31}{}{253--258}.
\PrintBackRefs{\CurrentBib}

\bibitem [\protect \citeauthoryear {%
Jacoby%
}{%
Jacoby%
}{%
{\protect \APACyear {2010}}%
}]{%
Jacoby2010}
\APACinsertmetastar {%
Jacoby2010}%
\begin{APACrefauthors}%
Jacoby, W\BPBI G.%
\end{APACrefauthors}%
\unskip\
\newblock
\APACrefYearMonthDay{2010}{}{}.
\newblock
{\BBOQ}\APACrefatitle {Policy attitudes, ideology and voting behavior in the
  2008 election} {Policy attitudes, ideology and voting behavior in the 2008
  election}.{\BBCQ}
\newblock
\APACjournalVolNumPages{Electoral Studies}{29}{}{557--568}.
\PrintBackRefs{\CurrentBib}

\bibitem [\protect \citeauthoryear {%
Judd%
\ \BBA {} Krosnick%
}{%
Judd%
\ \BBA {} Krosnick%
}{%
{\protect \APACyear {1989}}%
}]{%
Judd1989}
\APACinsertmetastar {%
Judd1989}%
\begin{APACrefauthors}%
Judd, C.%
\BCBT {}\ \BBA {} Krosnick, J.%
\end{APACrefauthors}%
\unskip\
\newblock
\APACrefYearMonthDay{1989}{}{}.
\newblock
{\BBOQ}\APACrefatitle {The structural bases of consistency among political
  attitudes: Effects of expertise and attitude importance} {The structural
  bases of consistency among political attitudes: Effects of expertise and
  attitude importance}.{\BBCQ}
\newblock
\BIn{} A.~Pratkanis, S.~Breckler\BCBL {}\ \BBA {} A.~Greenwald\ (\BEDS),
  \APACrefbtitle {Attitude structure and function} {Attitude structure and
  function}\ (\BPGS\ 99--128).
\newblock
\APACaddressPublisher{}{Hillsdale, NJ: Lawrence Erlbaum}.
\PrintBackRefs{\CurrentBib}

\bibitem [\protect \citeauthoryear {%
Judd%
, Krosnick%
\BCBL {}\ \BBA {} Milburn%
}{%
Judd%
\ \protect \BOthers {.}}{%
{\protect \APACyear {1980}}%
}]{%
Judd1980}
\APACinsertmetastar {%
Judd1980}%
\begin{APACrefauthors}%
Judd, C.%
, Krosnick, J.%
\BCBL {}\ \BBA {} Milburn, M.%
\end{APACrefauthors}%
\unskip\
\newblock
\APACrefYearMonthDay{1980}{}{}.
\newblock
{\BBOQ}\APACrefatitle {The structure of attitude systems in the general public:
  Comparisons of a structural equation model} {The structure of attitude
  systems in the general public: Comparisons of a structural equation
  model}.{\BBCQ}
\newblock
\APACjournalVolNumPages{American Sociological Review}{45}{}{627--643}.
\PrintBackRefs{\CurrentBib}

\bibitem [\protect \citeauthoryear {%
Judd%
, Krosnick%
\BCBL {}\ \BBA {} Milburn%
}{%
Judd%
\ \protect \BOthers {.}}{%
{\protect \APACyear {1981}}%
}]{%
Judd1981}
\APACinsertmetastar {%
Judd1981}%
\begin{APACrefauthors}%
Judd, C.%
, Krosnick, J.%
\BCBL {}\ \BBA {} Milburn, M.%
\end{APACrefauthors}%
\unskip\
\newblock
\APACrefYearMonthDay{1981}{}{}.
\newblock
{\BBOQ}\APACrefatitle {Political involvement and attitude structure in the
  general public} {Political involvement and attitude structure in the general
  public}.{\BBCQ}
\newblock
\APACjournalVolNumPages{American Sociological Review}{46}{}{660--669}.
\PrintBackRefs{\CurrentBib}

\bibitem [\protect \citeauthoryear {%
Kolaczyk%
}{%
Kolaczyk%
}{%
{\protect \APACyear {2009}}%
}]{%
Kolaczyk2009}
\APACinsertmetastar {%
Kolaczyk2009}%
\begin{APACrefauthors}%
Kolaczyk, E\BPBI D.%
\end{APACrefauthors}%
\unskip\
\newblock
\APACrefYear{2009}.
\newblock
\APACrefbtitle {Statistical Analysis of Network Data: Methods and Models}
  {Statistical analysis of network data: Methods and models}.
\newblock
\APACaddressPublisher{}{New York, NY: Springer}.
\PrintBackRefs{\CurrentBib}

\bibitem [\protect \citeauthoryear {%
Kraus%
}{%
Kraus%
}{%
{\protect \APACyear {1986}}%
}]{%
Kraus1986}
\APACinsertmetastar {%
Kraus1986}%
\begin{APACrefauthors}%
Kraus, S\BPBI J.%
\end{APACrefauthors}%
\unskip\
\newblock
\APACrefYearMonthDay{1986}{}{}.
\newblock
{\BBOQ}\APACrefatitle {Attitudes and the prediction of behavior: A
  meta-analysis of the empirical literature} {Attitudes and the prediction of
  behavior: A meta-analysis of the empirical literature}.{\BBCQ}
\newblock
\APACjournalVolNumPages{Personality and Social Psychology
  Bulletin}{51}{}{505--514}.
\PrintBackRefs{\CurrentBib}

\bibitem [\protect \citeauthoryear {%
Krosnick%
}{%
Krosnick%
}{%
{\protect \APACyear {1988}}%
}]{%
Krosnick1988}
\APACinsertmetastar {%
Krosnick1988}%
\begin{APACrefauthors}%
Krosnick, J\BPBI A.%
\end{APACrefauthors}%
\unskip\
\newblock
\APACrefYearMonthDay{1988}{}{}.
\newblock
{\BBOQ}\APACrefatitle {The role of attitude importance in social evaluation: a
  study of policy preferences, presidential candidate evaluations, and voting
  behavior} {The role of attitude importance in social evaluation: a study of
  policy preferences, presidential candidate evaluations, and voting
  behavior}.{\BBCQ}
\newblock
\APACjournalVolNumPages{Journal of Personality and Social
  Psychology}{55}{}{196--210}.
\PrintBackRefs{\CurrentBib}

\bibitem [\protect \citeauthoryear {%
Krosnick%
}{%
Krosnick%
}{%
{\protect \APACyear {1989}}%
}]{%
Krosnick1989}
\APACinsertmetastar {%
Krosnick1989}%
\begin{APACrefauthors}%
Krosnick, J\BPBI A.%
\end{APACrefauthors}%
\unskip\
\newblock
\APACrefYearMonthDay{1989}{}{}.
\newblock
{\BBOQ}\APACrefatitle {Attitude importance and attitude accessibility}
  {Attitude importance and attitude accessibility}.{\BBCQ}
\newblock
\APACjournalVolNumPages{Personality and Social Psychology
  Bulletin}{15}{}{297--308}.
\PrintBackRefs{\CurrentBib}

\bibitem [\protect \citeauthoryear {%
Krosnick%
, Boninger%
, Chuang%
, Berent%
\BCBL {}\ \BBA {} Carnot%
}{%
Krosnick%
\ \protect \BOthers {.}}{%
{\protect \APACyear {1993}}%
}]{%
Krosnick1993}
\APACinsertmetastar {%
Krosnick1993}%
\begin{APACrefauthors}%
Krosnick, J\BPBI A.%
, Boninger, D\BPBI S.%
, Chuang, Y\BPBI C.%
, Berent, M\BPBI K.%
\BCBL {}\ \BBA {} Carnot, C\BPBI G.%
\end{APACrefauthors}%
\unskip\
\newblock
\APACrefYearMonthDay{1993}{}{}.
\newblock
{\BBOQ}\APACrefatitle {Attitude strength: One construct or many related
  constructs?} {Attitude strength: One construct or many related
  constructs?}{\BBCQ}
\newblock
\APACjournalVolNumPages{Journal of Personality and Social
  Psychology}{65}{}{1132--1151}.
\PrintBackRefs{\CurrentBib}

\bibitem [\protect \citeauthoryear {%
Krosnick%
\ \BBA {} Petty%
}{%
Krosnick%
\ \BBA {} Petty%
}{%
{\protect \APACyear {1995}}%
}]{%
Krosnick1995}
\APACinsertmetastar {%
Krosnick1995}%
\begin{APACrefauthors}%
Krosnick, J\BPBI A.%
\BCBT {}\ \BBA {} Petty, R\BPBI E.%
\end{APACrefauthors}%
\unskip\
\newblock
\APACrefYearMonthDay{1995}{}{}.
\newblock
{\BBOQ}\APACrefatitle {Attitude strength: An overview} {Attitude strength: An
  overview}.{\BBCQ}
\newblock
\BIn{} R\BPBI E.~Petty\ \BBA {} J\BPBI A.~Krosnick\ (\BEDS), \APACrefbtitle
  {Attitude strength: Antecedents and consequences} {Attitude strength:
  Antecedents and consequences}\ (\BPGS\ 1--24).
\newblock
\APACaddressPublisher{}{Hillsdale, NJ: Lawrence Erlbaum}.
\PrintBackRefs{\CurrentBib}

\bibitem [\protect \citeauthoryear {%
Kruglanski%
\ \protect \BOthers {.}}{%
Kruglanski%
\ \protect \BOthers {.}}{%
{\protect \APACyear {2015}}%
}]{%
Kruglanski2015}
\APACinsertmetastar {%
Kruglanski2015}%
\begin{APACrefauthors}%
Kruglanski, A\BPBI W.%
, Katarzyna, J.%
, Chernikova, M.%
, Milyavsky, M.%
, Babush, M.%
, Baldner, C.%
\BCBL {}\ \BBA {} Pierro, A.%
\end{APACrefauthors}%
\unskip\
\newblock
\APACrefYearMonthDay{2015}{}{}.
\newblock
{\BBOQ}\APACrefatitle {The rocky road from attitudes to behaviors: Charting the
  goal systemic course of actions} {The rocky road from attitudes to behaviors:
  Charting the goal systemic course of actions}.{\BBCQ}
\newblock
\APACjournalVolNumPages{Psychological Review}{122}{}{598--620}.
\PrintBackRefs{\CurrentBib}

\bibitem [\protect \citeauthoryear {%
LaPiere%
}{%
LaPiere%
}{%
{\protect \APACyear {1934}}%
}]{%
LaPiere1934}
\APACinsertmetastar {%
LaPiere1934}%
\begin{APACrefauthors}%
LaPiere, E.%
\end{APACrefauthors}%
\unskip\
\newblock
\APACrefYearMonthDay{1934}{}{}.
\newblock
{\BBOQ}\APACrefatitle {Attitudes vs. actions} {Attitudes vs. actions}.{\BBCQ}
\newblock
\APACjournalVolNumPages{Social Forces}{13}{}{230--237}.
\PrintBackRefs{\CurrentBib}

\bibitem [\protect \citeauthoryear {%
Lavine%
, Thomsen%
, Zanna%
\BCBL {}\ \BBA {} Borgida%
}{%
Lavine%
\ \protect \BOthers {.}}{%
{\protect \APACyear {1998}}%
}]{%
Lavine1998}
\APACinsertmetastar {%
Lavine1998}%
\begin{APACrefauthors}%
Lavine, H.%
, Thomsen, C\BPBI J.%
, Zanna, M\BPBI P.%
\BCBL {}\ \BBA {} Borgida, E.%
\end{APACrefauthors}%
\unskip\
\newblock
\APACrefYearMonthDay{1998}{}{}.
\newblock
{\BBOQ}\APACrefatitle {On the primacy of affect in the determination of
  attitudes and behavior: The moderating role of affective-cognitive
  ambivalence} {On the primacy of affect in the determination of attitudes and
  behavior: The moderating role of affective-cognitive ambivalence}.{\BBCQ}
\newblock
\APACjournalVolNumPages{Journal of Experimental Social
  Psychology}{34}{}{398--421}.
\PrintBackRefs{\CurrentBib}

\bibitem [\protect \citeauthoryear {%
Little%
}{%
Little%
}{%
{\protect \APACyear {1988}}%
}]{%
Little1988}
\APACinsertmetastar {%
Little1988}%
\begin{APACrefauthors}%
Little, R\BPBI J\BPBI A.%
\end{APACrefauthors}%
\unskip\
\newblock
\APACrefYearMonthDay{1988}{}{}.
\newblock
{\BBOQ}\APACrefatitle {Missing-data adjustments in large surveys} {Missing-data
  adjustments in large surveys}.{\BBCQ}
\newblock
\APACjournalVolNumPages{Journal of Business \& Economic
  Statistics}{6}{}{287--296}.
\PrintBackRefs{\CurrentBib}

\bibitem [\protect \citeauthoryear {%
Markus%
}{%
Markus%
}{%
{\protect \APACyear {1982}}%
}]{%
Markus1982}
\APACinsertmetastar {%
Markus1982}%
\begin{APACrefauthors}%
Markus, G\BPBI B.%
\end{APACrefauthors}%
\unskip\
\newblock
\APACrefYearMonthDay{1982}{}{}.
\newblock
{\BBOQ}\APACrefatitle {Political attitudes during an election year: A report on
  the 1980 NES panel study} {Political attitudes during an election year: A
  report on the 1980 nes panel study}.{\BBCQ}
\newblock
\APACjournalVolNumPages{American Political Science Review}{76}{}{538--560}.
\PrintBackRefs{\CurrentBib}

\bibitem [\protect \citeauthoryear {%
Millar%
\ \BBA {} Millar%
}{%
Millar%
\ \BBA {} Millar%
}{%
{\protect \APACyear {1996}}%
}]{%
Millar1996}
\APACinsertmetastar {%
Millar1996}%
\begin{APACrefauthors}%
Millar, M\BPBI G.%
\BCBT {}\ \BBA {} Millar, K\BPBI U.%
\end{APACrefauthors}%
\unskip\
\newblock
\APACrefYearMonthDay{1996}{}{}.
\newblock
{\BBOQ}\APACrefatitle {Affective and Cognitive Responses and the
  Attitude-Behavior Relation} {Affective and cognitive responses and the
  attitude-behavior relation}.{\BBCQ}
\newblock
\APACjournalVolNumPages{Journal of Experimental Social
  Psychology}{32}{}{561--579}.
\PrintBackRefs{\CurrentBib}

\bibitem [\protect \citeauthoryear {%
Miller%
}{%
Miller%
}{%
{\protect \APACyear {1991}}%
}]{%
Miller1991}
\APACinsertmetastar {%
Miller1991}%
\begin{APACrefauthors}%
Miller, W\BPBI E.%
\end{APACrefauthors}%
\unskip\
\newblock
\APACrefYearMonthDay{1991}{}{}.
\newblock
{\BBOQ}\APACrefatitle {Party identification, realignment , and party voting:
  Back to the basics} {Party identification, realignment , and party voting:
  Back to the basics}.{\BBCQ}
\newblock
\APACjournalVolNumPages{American Political Science Review}{85}{}{557--568}.
\PrintBackRefs{\CurrentBib}

\bibitem [\protect \citeauthoryear {%
Monroe%
\ \BBA {} Read%
}{%
Monroe%
\ \BBA {} Read%
}{%
{\protect \APACyear {2008}}%
}]{%
Monroe2008}
\APACinsertmetastar {%
Monroe2008}%
\begin{APACrefauthors}%
Monroe, B\BPBI M.%
\BCBT {}\ \BBA {} Read, S\BPBI J.%
\end{APACrefauthors}%
\unskip\
\newblock
\APACrefYearMonthDay{2008}{}{}.
\newblock
{\BBOQ}\APACrefatitle {{A general connectionist model of attitude structure and
  change: the ACS (Attitudes as Constraint Satisfaction) model}} {{A general
  connectionist model of attitude structure and change: the ACS (Attitudes as
  Constraint Satisfaction) model}}.{\BBCQ}
\newblock
\APACjournalVolNumPages{Psychological Review}{115}{}{733--759}.
\PrintBackRefs{\CurrentBib}

\bibitem [\protect \citeauthoryear {%
Murphy%
}{%
Murphy%
}{%
{\protect \APACyear {2012}}%
}]{%
Murphy2012}
\APACinsertmetastar {%
Murphy2012}%
\begin{APACrefauthors}%
Murphy, K.%
\end{APACrefauthors}%
\unskip\
\newblock
\APACrefYear{2012}.
\newblock
\APACrefbtitle {Machine learning: A probabilistic perspective} {Machine
  learning: A probabilistic perspective}.
\newblock
\APACaddressPublisher{}{Cambridge, MA: MIT Press}.
\PrintBackRefs{\CurrentBib}

\bibitem [\protect \citeauthoryear {%
Nadeau%
\ \BBA {} Lewis-Beck%
}{%
Nadeau%
\ \BBA {} Lewis-Beck%
}{%
{\protect \APACyear {2001}}%
}]{%
Nadeau2001}
\APACinsertmetastar {%
Nadeau2001}%
\begin{APACrefauthors}%
Nadeau, R.%
\BCBT {}\ \BBA {} Lewis-Beck, M\BPBI S.%
\end{APACrefauthors}%
\unskip\
\newblock
\APACrefYearMonthDay{2001}{}{}.
\newblock
{\BBOQ}\APACrefatitle {{National economic voting in U.S. presidential
  elections}} {{National economic voting in U.S. presidential
  elections}}.{\BBCQ}
\newblock
\APACjournalVolNumPages{Journal of Politics}{63}{}{159--181}.
\PrintBackRefs{\CurrentBib}

\bibitem [\protect \citeauthoryear {%
Newman%
}{%
Newman%
}{%
{\protect \APACyear {2010}}%
}]{%
Newman2010}
\APACinsertmetastar {%
Newman2010}%
\begin{APACrefauthors}%
Newman, M\BPBI E\BPBI J.%
\end{APACrefauthors}%
\unskip\
\newblock
\APACrefYear{2010}.
\newblock
\APACrefbtitle {Networks: An Introduction} {Networks: An introduction}.
\newblock
\APACaddressPublisher{}{Oxford: Oxford University Press}.
\PrintBackRefs{\CurrentBib}

\bibitem [\protect \citeauthoryear {%
Opsahl%
, Agneessens%
\BCBL {}\ \BBA {} Skvoretz%
}{%
Opsahl%
\ \protect \BOthers {.}}{%
{\protect \APACyear {2010}}%
}]{%
Opsahl2010}
\APACinsertmetastar {%
Opsahl2010}%
\begin{APACrefauthors}%
Opsahl, T.%
, Agneessens, F.%
\BCBL {}\ \BBA {} Skvoretz, J.%
\end{APACrefauthors}%
\unskip\
\newblock
\APACrefYearMonthDay{2010}{}{}.
\newblock
{\BBOQ}\APACrefatitle {Node centrality in weighted networks: Generalizing
  degree and shortest paths} {Node centrality in weighted networks:
  Generalizing degree and shortest paths}.{\BBCQ}
\newblock
\APACjournalVolNumPages{Social Networks}{32}{}{245--251}.
\PrintBackRefs{\CurrentBib}

\bibitem [\protect \citeauthoryear {%
Palfrey%
\ \BBA {} Poole%
}{%
Palfrey%
\ \BBA {} Poole%
}{%
{\protect \APACyear {1978}}%
}]{%
Fiorina1978}
\APACinsertmetastar {%
Fiorina1978}%
\begin{APACrefauthors}%
Palfrey, T\BPBI R.%
\BCBT {}\ \BBA {} Poole, K\BPBI T.%
\end{APACrefauthors}%
\unskip\
\newblock
\APACrefYearMonthDay{1978}{}{}.
\newblock
{\BBOQ}\APACrefatitle {{Economic retrospective voting in American national
  elections: A micro-analysis}} {{Economic retrospective voting in American
  national elections: A micro-analysis}}.{\BBCQ}
\newblock
\APACjournalVolNumPages{American Journal of Political Science}{22}{}{426--443}.
\PrintBackRefs{\CurrentBib}

\bibitem [\protect \citeauthoryear {%
Palfrey%
\ \BBA {} Poole%
}{%
Palfrey%
\ \BBA {} Poole%
}{%
{\protect \APACyear {1987}}%
}]{%
Palfrey1987}
\APACinsertmetastar {%
Palfrey1987}%
\begin{APACrefauthors}%
Palfrey, T\BPBI R.%
\BCBT {}\ \BBA {} Poole, K\BPBI T.%
\end{APACrefauthors}%
\unskip\
\newblock
\APACrefYearMonthDay{1987}{}{}.
\newblock
{\BBOQ}\APACrefatitle {The relationship between information, ideology, and
  voting behavior} {The relationship between information, ideology, and voting
  behavior}.{\BBCQ}
\newblock
\APACjournalVolNumPages{American Journal of Political Science}{31}{}{511--530}.
\PrintBackRefs{\CurrentBib}

\bibitem [\protect \citeauthoryear {%
Rabinowitz%
\ \BBA {} MacDonald%
}{%
Rabinowitz%
\ \BBA {} MacDonald%
}{%
{\protect \APACyear {1989}}%
}]{%
Rabinowitz1989}
\APACinsertmetastar {%
Rabinowitz1989}%
\begin{APACrefauthors}%
Rabinowitz, G.%
\BCBT {}\ \BBA {} MacDonald, S\BPBI E.%
\end{APACrefauthors}%
\unskip\
\newblock
\APACrefYearMonthDay{1989}{}{}.
\newblock
{\BBOQ}\APACrefatitle {A directional theory of issue voting} {A directional
  theory of issue voting}.{\BBCQ}
\newblock
\APACjournalVolNumPages{American Political Science Review}{83}{}{93--121}.
\PrintBackRefs{\CurrentBib}

\bibitem [\protect \citeauthoryear {%
Rosenberg%
, Hovland%
, McGuire%
\BCBL {}\ \BBA {} Brehm%
}{%
Rosenberg%
\ \protect \BOthers {.}}{%
{\protect \APACyear {1960}}%
}]{%
Rosenberg1960}
\APACinsertmetastar {%
Rosenberg1960}%
\begin{APACrefauthors}%
Rosenberg, M\BPBI J.%
, Hovland, C\BPBI I.%
, McGuire, R\BPBI P., W. J.and~Abelson%
\BCBL {}\ \BBA {} Brehm, J\BPBI W.%
\end{APACrefauthors}%
\unskip\
\newblock
\APACrefYear{1960}.
\newblock
\APACrefbtitle {Attitude Organization and Change: An Analysis of Consistency
  among Attitude Components} {Attitude organization and change: An analysis of
  consistency among attitude components}.
\newblock
\APACaddressPublisher{}{New Haven, CA: Yale University Press}.
\PrintBackRefs{\CurrentBib}

\bibitem [\protect \citeauthoryear {%
Taleb%
}{%
Taleb%
}{%
{\protect \APACyear {2017}}%
}]{%
Taleb2017}
\APACinsertmetastar {%
Taleb2017}%
\begin{APACrefauthors}%
Taleb, N.%
\end{APACrefauthors}%
\unskip\
\newblock
\APACrefYearMonthDay{2017}{}{}.
\newblock
\APACrefbtitle {How to forecast an election.} {How to forecast an election.}
\newblock
\APACrefnote{https://arxiv.org/abs/1703.06351}
\PrintBackRefs{\CurrentBib}

\bibitem [\protect \citeauthoryear {%
Thompson%
, Zanna%
\BCBL {}\ \BBA {} Griffin%
}{%
Thompson%
\ \protect \BOthers {.}}{%
{\protect \APACyear {1995}}%
}]{%
Thompson1995}
\APACinsertmetastar {%
Thompson1995}%
\begin{APACrefauthors}%
Thompson, M\BPBI M.%
, Zanna, M\BPBI P.%
\BCBL {}\ \BBA {} Griffin, D\BPBI W.%
\end{APACrefauthors}%
\unskip\
\newblock
\APACrefYearMonthDay{1995}{}{}.
\newblock
{\BBOQ}\APACrefatitle {Let's not be indifferent about (attitudinal)
  ambivalence} {Let's not be indifferent about (attitudinal)
  ambivalence}.{\BBCQ}
\newblock
\BIn{} R\BPBI E.~Petty\ \BBA {} J\BPBI A.~Krosnick\ (\BEDS), \APACrefbtitle
  {Attitude strength: Antecedents and consequences} {Attitude strength:
  Antecedents and consequences}\ (\BPGS\ 361--386).
\newblock
\APACaddressPublisher{}{Hillsdale, NJ: Lawrence Erlbaum}.
\PrintBackRefs{\CurrentBib}

\bibitem [\protect \citeauthoryear {%
van Borkulo%
\ \protect \BOthers {.}}{%
van Borkulo%
\ \protect \BOthers {.}}{%
{\protect \APACyear {2014}}%
}]{%
vanBorkulo2014}
\APACinsertmetastar {%
vanBorkulo2014}%
\begin{APACrefauthors}%
van Borkulo, C\BPBI D.%
, Borsboom, D.%
, Epskamp, S.%
, Blanken, B\BPBI W.%
, Bosschloo, L.%
, Schoevers, R\BPBI A.%
\BCBL {}\ \BBA {} Waldorp, L\BPBI J.%
\end{APACrefauthors}%
\unskip\
\newblock
\APACrefYearMonthDay{2014}{}{}.
\newblock
{\BBOQ}\APACrefatitle {A new method for constructing networks from binary data}
  {A new method for constructing networks from binary data}.{\BBCQ}
\newblock
\APACjournalVolNumPages{Scientific Reports}{4}{}{5918}.
\PrintBackRefs{\CurrentBib}

\bibitem [\protect \citeauthoryear {%
van Borkulo%
\ \protect \BOthers {.}}{%
van Borkulo%
\ \protect \BOthers {.}}{%
{\protect \APACyear {2015}}%
}]{%
vanBorkulo2015}
\APACinsertmetastar {%
vanBorkulo2015}%
\begin{APACrefauthors}%
van Borkulo, C\BPBI D.%
, Boschloo, L.%
, Borsboom, D.%
, Penninx, B\BPBI W\BPBI J\BPBI H.%
, Waldorp, L\BPBI J.%
\BCBL {}\ \BBA {} Schoevers, R\BPBI A.%
\end{APACrefauthors}%
\unskip\
\newblock
\APACrefYearMonthDay{2015}{}{}.
\newblock
{\BBOQ}\APACrefatitle {Association of symptom network structure with the course
  of depression} {Association of symptom network structure with the course of
  depression}.{\BBCQ}
\newblock
\APACjournalVolNumPages{JAMA Psychiatry}{72}{}{1219--1226}.
\PrintBackRefs{\CurrentBib}

\bibitem [\protect \citeauthoryear {%
van Buuren%
\ \BBA {} Groothuis-Oudshoorn%
}{%
van Buuren%
\ \BBA {} Groothuis-Oudshoorn%
}{%
{\protect \APACyear {2011}}%
}]{%
vanBuuren2011}
\APACinsertmetastar {%
vanBuuren2011}%
\begin{APACrefauthors}%
van Buuren, S.%
\BCBT {}\ \BBA {} Groothuis-Oudshoorn, K.%
\end{APACrefauthors}%
\unskip\
\newblock
\APACrefYearMonthDay{2011}{}{}.
\newblock
{\BBOQ}\APACrefatitle {Multivariate imputation by chained equations}
  {Multivariate imputation by chained equations}.{\BBCQ}
\newblock
\APACjournalVolNumPages{Journal of Statistical Software}{45}{}{1--67}.
\PrintBackRefs{\CurrentBib}

\bibitem [\protect \citeauthoryear {%
van~de Leemput%
\ \protect \BOthers {.}}{%
van~de Leemput%
\ \protect \BOthers {.}}{%
{\protect \APACyear {2014}}%
}]{%
vandeLemmput2014}
\APACinsertmetastar {%
vandeLemmput2014}%
\begin{APACrefauthors}%
van~de Leemput, I\BPBI A.%
, Wichers, M.%
, Cramer, A\BPBI O\BPBI J.%
, Borsboom, D.%
, Tuerlinckx, F.%
, Kuppens, P.%
\BDBL {}Scheffer, M.%
\end{APACrefauthors}%
\unskip\
\newblock
\APACrefYearMonthDay{2014}{}{}.
\newblock
{\BBOQ}\APACrefatitle {Critical slowing down as early warning for the onset and
  termination of depression} {Critical slowing down as early warning for the
  onset and termination of depression}.{\BBCQ}
\newblock
\APACjournalVolNumPages{Proceedings of the National Academy of Sciences of the
  United States of America}{111}{}{87--92}.
\PrintBackRefs{\CurrentBib}

\bibitem [\protect \citeauthoryear {%
van~den Berg%
, Manstead%
, van~der Pligt%
\BCBL {}\ \BBA {} Wigboldus%
}{%
van~den Berg%
\ \protect \BOthers {.}}{%
{\protect \APACyear {2005}}%
}]{%
vandenBerg2005}
\APACinsertmetastar {%
vandenBerg2005}%
\begin{APACrefauthors}%
van~den Berg, H.%
, Manstead, A\BPBI S\BPBI R.%
, van~der Pligt, J.%
\BCBL {}\ \BBA {} Wigboldus, D\BPBI H\BPBI J.%
\end{APACrefauthors}%
\unskip\
\newblock
\APACrefYearMonthDay{2005}{}{}.
\newblock
{\BBOQ}\APACrefatitle {The role of affect in attitudes toward organ donation
  and donor-relevant decisions} {The role of affect in attitudes toward organ
  donation and donor-relevant decisions}.{\BBCQ}
\newblock
\APACjournalVolNumPages{Psychology and Health}{20}{}{789--802}.
\PrintBackRefs{\CurrentBib}

\bibitem [\protect \citeauthoryear {%
van~der Pligt%
, de Vries%
, Manstead%
\BCBL {}\ \BBA {} van Harreveld%
}{%
van~der Pligt%
\ \protect \BOthers {.}}{%
{\protect \APACyear {2000}}%
}]{%
vanderPligt2000}
\APACinsertmetastar {%
vanderPligt2000}%
\begin{APACrefauthors}%
van~der Pligt, J.%
, de Vries, N\BPBI K.%
, Manstead, A\BPBI S\BPBI R.%
\BCBL {}\ \BBA {} van Harreveld, F.%
\end{APACrefauthors}%
\unskip\
\newblock
\APACrefYearMonthDay{2000}{}{}.
\newblock
{\BBOQ}\APACrefatitle {The importance of being selective: Weighing the role of
  attribute importance} {The importance of being selective: Weighing the role
  of attribute importance}.{\BBCQ}
\newblock
\APACjournalVolNumPages{Advances in Experimental Social
  Psychology}{32}{}{135--200}.
\PrintBackRefs{\CurrentBib}

\bibitem [\protect \citeauthoryear {%
van Harreveld%
, Nohlen%
\BCBL {}\ \BBA {} Schneider%
}{%
van Harreveld%
\ \protect \BOthers {.}}{%
{\protect \APACyear {2015}}%
}]{%
vanHarreveld2015}
\APACinsertmetastar {%
vanHarreveld2015}%
\begin{APACrefauthors}%
van Harreveld, F.%
, Nohlen, H\BPBI U.%
\BCBL {}\ \BBA {} Schneider, I\BPBI K.%
\end{APACrefauthors}%
\unskip\
\newblock
\APACrefYearMonthDay{2015}{}{}.
\newblock
{\BBOQ}\APACrefatitle {The ABC of ambivalence: Affective, behavioral, and
  cognitive consequences of attitudinal conflict} {The abc of ambivalence:
  Affective, behavioral, and cognitive consequences of attitudinal
  conflict}.{\BBCQ}
\newblock
\APACjournalVolNumPages{Advances in Experimental Social
  Psychology}{52}{}{285--324}.
\PrintBackRefs{\CurrentBib}

\bibitem [\protect \citeauthoryear {%
van Harreveld%
, van~der Pligt%
, de Vries%
\BCBL {}\ \BBA {} Andreas%
}{%
van Harreveld%
\ \protect \BOthers {.}}{%
{\protect \APACyear {2000}}%
}]{%
vanHarreveld2000}
\APACinsertmetastar {%
vanHarreveld2000}%
\begin{APACrefauthors}%
van Harreveld, F.%
, van~der Pligt, J.%
, de Vries, N\BPBI K.%
\BCBL {}\ \BBA {} Andreas, S.%
\end{APACrefauthors}%
\unskip\
\newblock
\APACrefYearMonthDay{2000}{}{}.
\newblock
{\BBOQ}\APACrefatitle {The structure of attitudes: attribute importance,
  accessibility and judgment} {The structure of attitudes: attribute
  importance, accessibility and judgment}.{\BBCQ}
\newblock
\APACjournalVolNumPages{British Journal of Social Psychology}{39}{}{363--380}.
\PrintBackRefs{\CurrentBib}

\bibitem [\protect \citeauthoryear {%
Visser%
, Bizer%
\BCBL {}\ \BBA {} Krosnick%
}{%
Visser%
\ \protect \BOthers {.}}{%
{\protect \APACyear {2006}}%
}]{%
Visser2006}
\APACinsertmetastar {%
Visser2006}%
\begin{APACrefauthors}%
Visser, P\BPBI S.%
, Bizer, G.%
\BCBL {}\ \BBA {} Krosnick, J\BPBI A.%
\end{APACrefauthors}%
\unskip\
\newblock
\APACrefYearMonthDay{2006}{}{}.
\newblock
{\BBOQ}\APACrefatitle {Exploring the latent structure of strength-related
  attitude attributes} {Exploring the latent structure of strength-related
  attitude attributes}.{\BBCQ}
\newblock
\APACjournalVolNumPages{Advances in Experimental Social
  Psychology}{38}{}{1--67}.
\PrintBackRefs{\CurrentBib}

\bibitem [\protect \citeauthoryear {%
Visser%
, Krosnick%
\BCBL {}\ \BBA {} Simmons%
}{%
Visser%
\ \protect \BOthers {.}}{%
{\protect \APACyear {2003}}%
}]{%
Visser2003}
\APACinsertmetastar {%
Visser2003}%
\begin{APACrefauthors}%
Visser, P\BPBI S.%
, Krosnick, J\BPBI A.%
\BCBL {}\ \BBA {} Simmons, J\BPBI P.%
\end{APACrefauthors}%
\unskip\
\newblock
\APACrefYearMonthDay{2003}{}{}.
\newblock
{\BBOQ}\APACrefatitle {Distinguishing the cognitive and behavioral consequences
  of attitude importance and certainty: A new approach to testing the
  common-factor hypothesis} {Distinguishing the cognitive and behavioral
  consequences of attitude importance and certainty: A new approach to testing
  the common-factor hypothesis}.{\BBCQ}
\newblock
\APACjournalVolNumPages{Journal of Experimental Social
  Psychology}{39}{}{118--141}.
\PrintBackRefs{\CurrentBib}

\bibitem [\protect \citeauthoryear {%
Wainwright%
\ \BBA {} Jordan%
}{%
Wainwright%
\ \BBA {} Jordan%
}{%
{\protect \APACyear {2008}}%
}]{%
Wainwright2008}
\APACinsertmetastar {%
Wainwright2008}%
\begin{APACrefauthors}%
Wainwright, M\BPBI J.%
\BCBT {}\ \BBA {} Jordan, M\BPBI I.%
\end{APACrefauthors}%
\unskip\
\newblock
\APACrefYearMonthDay{2008}{}{}.
\newblock
{\BBOQ}\APACrefatitle {Graphical models, exponential families, and variational
  inference} {Graphical models, exponential families, and variational
  inference}.{\BBCQ}
\newblock
\APACjournalVolNumPages{Foundations and Trends\textregistered in Machine
  Learning}{1}{}{1--305}.
\PrintBackRefs{\CurrentBib}

\bibitem [\protect \citeauthoryear {%
Wallis%
}{%
Wallis%
}{%
{\protect \APACyear {2007}}%
}]{%
Wallis2007}
\APACinsertmetastar {%
Wallis2007}%
\begin{APACrefauthors}%
Wallis, W\BPBI D.%
\end{APACrefauthors}%
\unskip\
\newblock
\APACrefYear{2007}.
\newblock
\APACrefbtitle {A Beginner's Guide to Graph Theory} {A beginner's guide to
  graph theory}.
\newblock
\APACaddressPublisher{}{New York, NY: Birkh\"auser}.
\PrintBackRefs{\CurrentBib}

\bibitem [\protect \citeauthoryear {%
Watts%
\ \BBA {} Strogatz%
}{%
Watts%
\ \BBA {} Strogatz%
}{%
{\protect \APACyear {1998}}%
}]{%
Watts1998}
\APACinsertmetastar {%
Watts1998}%
\begin{APACrefauthors}%
Watts, D\BPBI J.%
\BCBT {}\ \BBA {} Strogatz, S\BPBI H.%
\end{APACrefauthors}%
\unskip\
\newblock
\APACrefYearMonthDay{1998}{}{}.
\newblock
{\BBOQ}\APACrefatitle {Collective dynamics of `small-world' networks}
  {Collective dynamics of `small-world' networks}.{\BBCQ}
\newblock
\APACjournalVolNumPages{Nature}{393}{}{440--442}.
\PrintBackRefs{\CurrentBib}

\bibitem [\protect \citeauthoryear {%
West%
}{%
West%
}{%
{\protect \APACyear {1996}}%
}]{%
West1996}
\APACinsertmetastar {%
West1996}%
\begin{APACrefauthors}%
West, D\BPBI B.%
\end{APACrefauthors}%
\unskip\
\newblock
\APACrefYear{1996}.
\newblock
\APACrefbtitle {Introduction to Graph Theory} {Introduction to graph theory}.
\newblock
\APACaddressPublisher{}{Upper Saddle River, NJ: Prentice Hall}.
\PrintBackRefs{\CurrentBib}

\bibitem [\protect \citeauthoryear {%
Wicker%
}{%
Wicker%
}{%
{\protect \APACyear {1969}}%
}]{%
Wicker1969}
\APACinsertmetastar {%
Wicker1969}%
\begin{APACrefauthors}%
Wicker, A\BPBI W.%
\end{APACrefauthors}%
\unskip\
\newblock
\APACrefYearMonthDay{1969}{}{}.
\newblock
{\BBOQ}\APACrefatitle {Attitudes versus actions: The relationship of verbal and
  overt behavioral responses to attitude objects} {Attitudes versus actions:
  The relationship of verbal and overt behavioral responses to attitude
  objects}.{\BBCQ}
\newblock
\APACjournalVolNumPages{Journal of Social Issues}{25}{}{41--78}.
\PrintBackRefs{\CurrentBib}

\bibitem [\protect \citeauthoryear {%
Wlezien%
}{%
Wlezien%
}{%
{\protect \APACyear {2003}}%
}]{%
Wlezien2003}
\APACinsertmetastar {%
Wlezien2003}%
\begin{APACrefauthors}%
Wlezien, C.%
\end{APACrefauthors}%
\unskip\
\newblock
\APACrefYearMonthDay{2003}{}{}.
\newblock
{\BBOQ}\APACrefatitle {Presidential Election Polls in 2000: A Study in
  Dynamics} {Presidential election polls in 2000: A study in dynamics}.{\BBCQ}
\newblock
\APACjournalVolNumPages{Presidential Studies Quarterly}{33}{}{172--186}.
\PrintBackRefs{\CurrentBib}

\bibitem [\protect \citeauthoryear {%
Young%
\ \BBA {} Smith%
}{%
Young%
\ \BBA {} Smith%
}{%
{\protect \APACyear {2005}}%
}]{%
Young2005}
\APACinsertmetastar {%
Young2005}%
\begin{APACrefauthors}%
Young, G.%
\BCBT {}\ \BBA {} Smith, R.%
\end{APACrefauthors}%
\unskip\
\newblock
\APACrefYear{2005}.
\newblock
\APACrefbtitle {Essentials of statistical inference} {Essentials of statistical
  inference}.
\newblock
\APACaddressPublisher{}{Cambridge, UK: Cambridge University Press}.
\PrintBackRefs{\CurrentBib}

\bibitem [\protect \citeauthoryear {%
Zanna%
\ \BBA {} Rempel%
}{%
Zanna%
\ \BBA {} Rempel%
}{%
{\protect \APACyear {1988}}%
}]{%
Zanna1988}
\APACinsertmetastar {%
Zanna1988}%
\begin{APACrefauthors}%
Zanna, M\BPBI P.%
\BCBT {}\ \BBA {} Rempel, J\BPBI K.%
\end{APACrefauthors}%
\unskip\
\newblock
\APACrefYearMonthDay{1988}{}{}.
\newblock
{\BBOQ}\APACrefatitle {Attitudes: A new look at an old concept} {Attitudes: A
  new look at an old concept}.{\BBCQ}
\newblock
\BIn{} D.~Bar-Tal\ \BBA {} A\BPBI W.~Kruglanski\ (\BEDS), \APACrefbtitle {The
  Psychology of Knowledge} {The psychology of knowledge}\ (\BPGS\ 315--334).
\newblock
\APACaddressPublisher{}{Cambridge, UK: Cambridge University Press}.
\PrintBackRefs{\CurrentBib}

\end{thebibliography}
\bibliographystyle{apacite}

\subsubsection*{Acknowledgements} 
We thank S. Epskamp and G. Costantini for help with the simulation studies; M. Deserno for help with the data analyses. D. B. was supported by a Consolidator Grant No. 647209 from the European Research Council.
\subsubsection*{Author Contributions} 
J.D. developed the study concept; J.D., D.B., F.v.H., and H.L.J.v.d.M contributed to the study design; J. D. performed the data analysis and interpretation under the supervision of D.B., F.v.H., and H.L.J.v.d.M.; J.D. drafted the manuscript, and D.B., F.v.H., and H.L.J.v.d.M. provided critical revisions. L.J.W. provided the analytical solutions of the hypotheses.
\subsubsection*{Author Information} 
Data used in this paper is available at www.electionstudies.org. Correspondence and request for materials should be addressed to J.D. (j.dalege@uva.nl).
\subsubsection*{Competing Financial Interest} 
The authors declare no competing financial interests.

\end{multicols}

\titleformat*{\section}{\normalfont\Large\bfseries}
\titleformat*{\subsection}{\normalfont\large\bfseries}
\titleformat*{\subsubsection}{\normalfont\bfseries}

\setcounter{table}{0}
\setcounter{figure}{0}
\setcounter{equation}{0}

\renewcommand{\figurename}{Supplementary Figure}
\renewcommand{\tablename}{Supplementary Table}

\title{\vspace{-2.0cm}\textbf{Supplementary Materials for: Network Structure Explains the Impact of Attitudes on Voting Decisions}}
\maketitle

\section*{Supplementary Figures}

\begin{figure}[!h]
\centering
\includegraphics[width=5.5in]{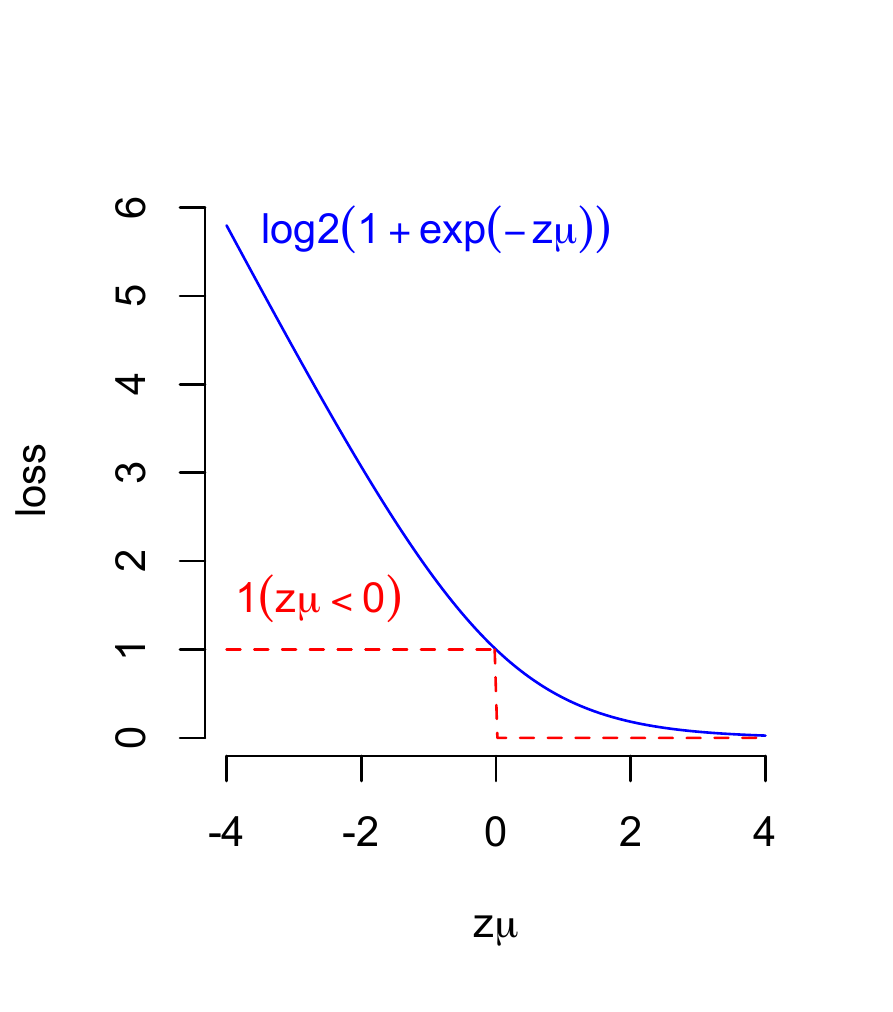}
\caption{\textbf{Misclassification loss $\mathcal{C}$ in (\ref{eq:mis-loss}) and $\psi$ in (\ref{eq:log-loss}) as a function of the margin $x\mu=(2y-1)\mu$. }}
\label{fig:figS1}
\end{figure}

\newpage

\begin{figure*}[!h]
    \centering
\includegraphics[width=6.27in]{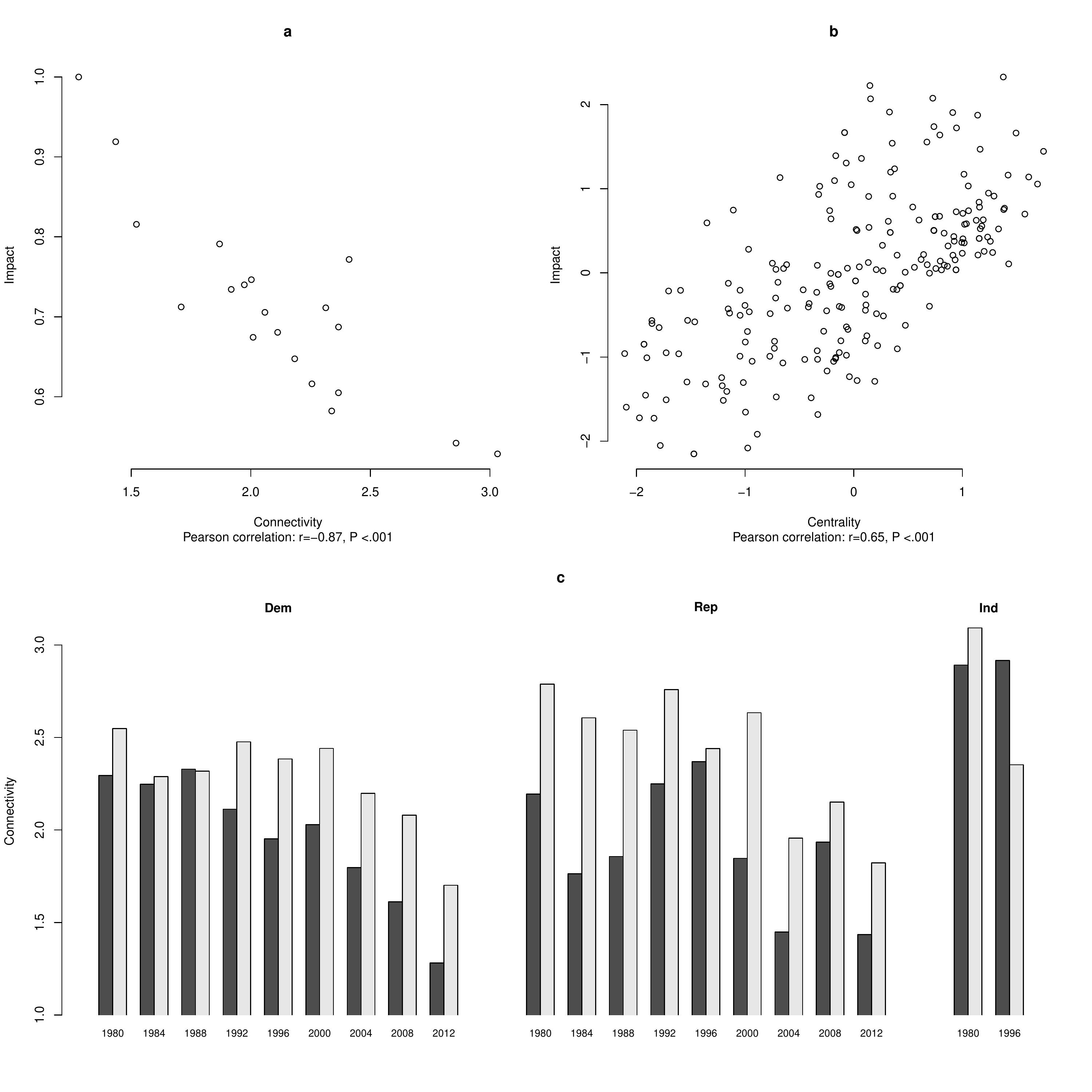}
\caption{\textbf{Analyses including non-voters.} (\textbf{a}) Relation between connectivity and global impact. (\textbf{b}) Relation between centrality and specific impact. (\textbf{c}) Comparison between connectivity of voters (dark grey bars) and non-voters (light grey bars). Dem: Democratic candidates; Rep: Republican candidates, Ind: Independent candidates.}
\label{fig:figS2}
\end{figure*}

\newpage

\begin{figure*}[!h]
    \centering
\includegraphics[width=6.27in]{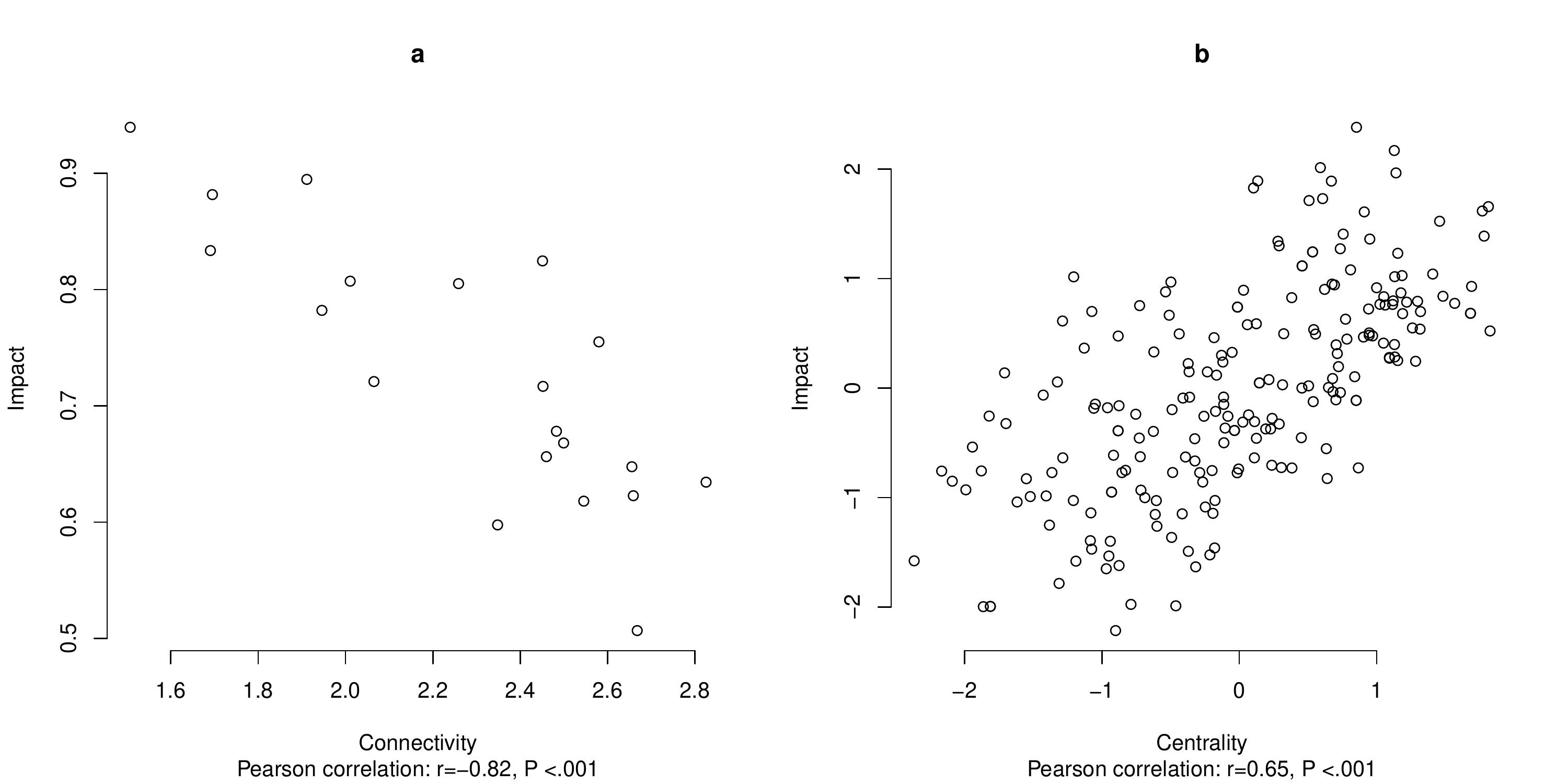}
\caption{\textbf{Analyses including only independents} (\textbf{a}) Relation between connectivity and global impact. (\textbf{b}) Relation between centrality and specific impact.}
\label{fig:figS3}
\end{figure*}

\newpage

\begin{figure*}[!h]
    \centering
\includegraphics[width=6.27in]{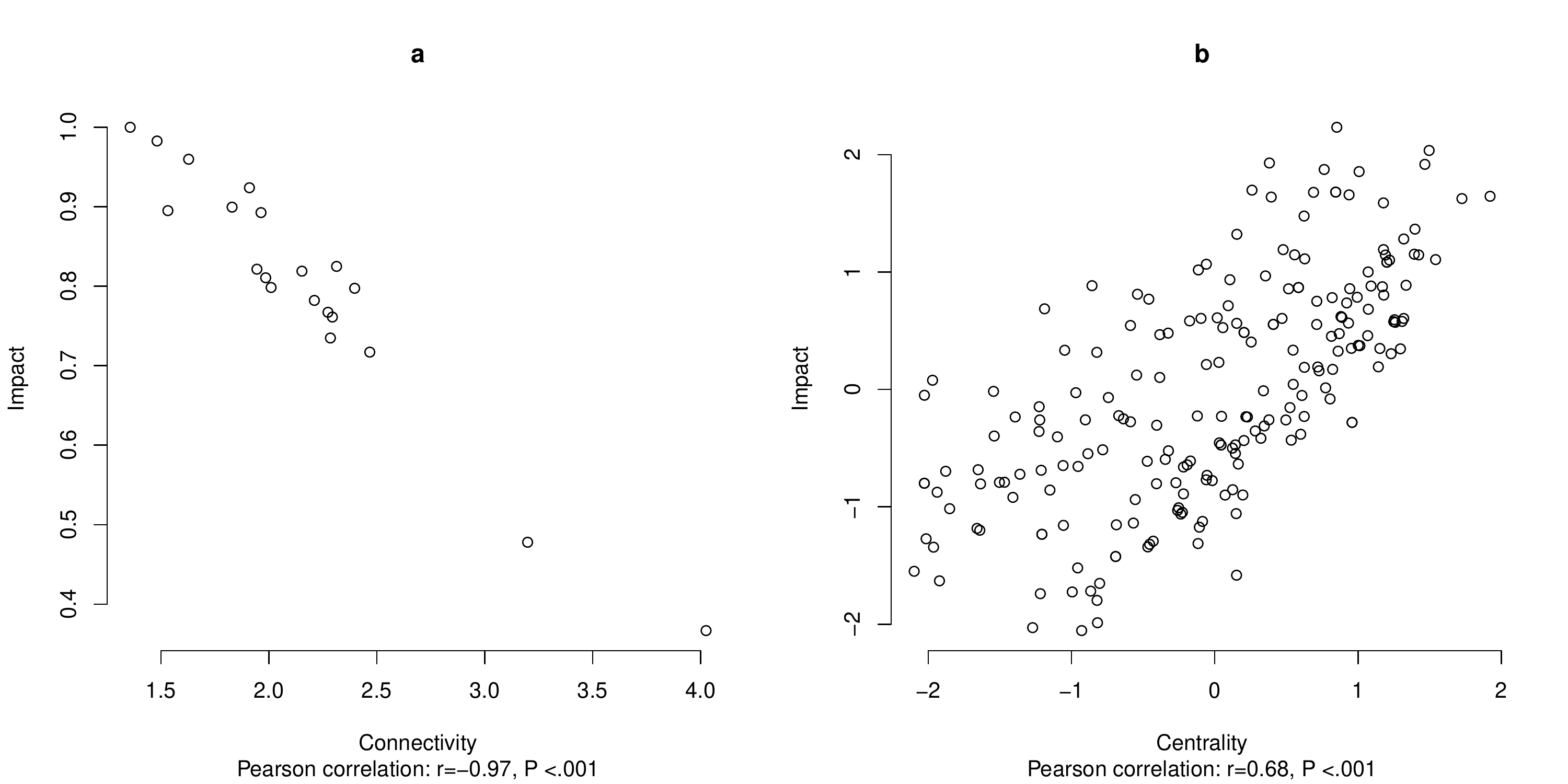}
\caption{\textbf{Analyses with imputed missing values} (\textbf{a}) Relation between connectivity and global impact. (\textbf{b}) Relation between centrality and specific impact.}
\label{fig:figS4}
\end{figure*}

\newpage

\begin{figure*}[!h]
    \centering
\includegraphics[width=6.27in]{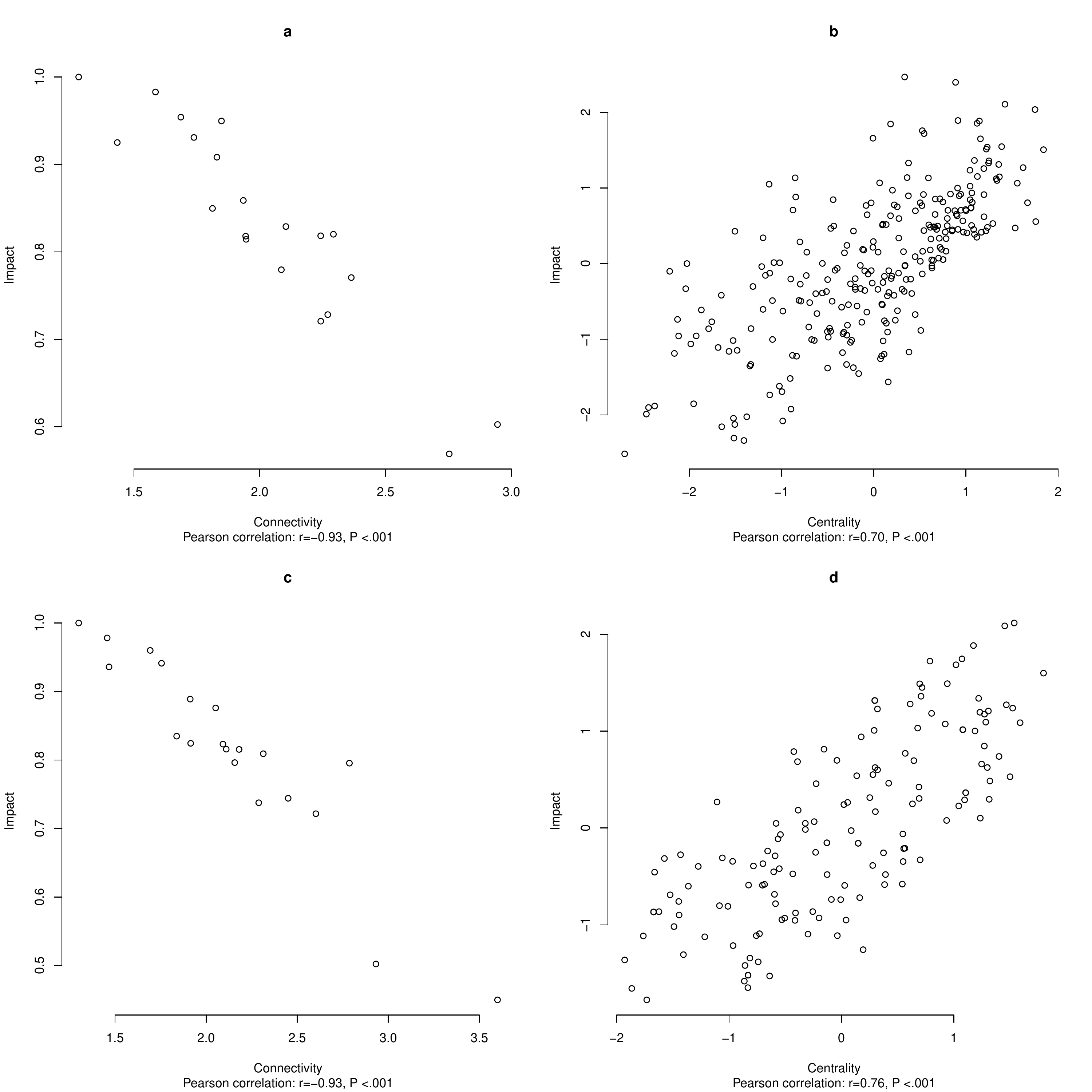}
\caption{\textbf{Analyses on different numbers of attitude elements} (\textbf{a}) Relation between connectivity and global impact for the analysis including all attitude elements. (\textbf{b}) Relation between centrality and specific impact for the analysis including all attitude elements. (\textbf{c}) Relation between connectivity and global impact for the analysis including seven attitude elements. (\textbf{d}) Relation between centrality and specific impact for the analysis including seven attitude elements. }
\label{fig:figS5}
\end{figure*}

\newpage
\subsection*{\color{black}{Supplementary Tables}}
\begin{table*}[h]
\caption{\textbf{Results of the Simulations for Each Network-Generating Algorithm and Edge Weight Distribution.}}
\label{tab:tabS1}
\begin{tabular}{llll}\hline
           \cline{1-4}
&Preferential attachment  & Small-world   &  Random graph  \\ \hline
&\multicolumn{3}{l}{\underline{Connectivity/impact correlations}}\\
Normal distribution & mean \textit{r}=-0.91 & mean \textit{r}=-0.91 & mean \textit{r}=-0.90\\
& s.d. \textit{r}=0.07 & s.d. \textit{r}=0.05 & s.d. \textit{r}=0.08\\
Power-law distribution& mean \textit{r}=-0.92 & mean \textit{r}=-0.91 & mean \textit{r}=-0.91\\
 & s.d. \textit{r}=0.05 & s.d. \textit{r}=0.05 & s.d. \textit{r}=0.04\\
Uniform distribution& mean \textit{r}=-0.92 & mean \textit{r}=-0.89 & mean \textit{r}=-0.90\\
 & s.d. \textit{r}=0.05 & s.d. \textit{r}=0.08 & s.d. \textit{r}=0.08\\
&\multicolumn{3}{l}{\underline{Centrality/impact correlations}}\\
Normal distribution & mean \textit{r}=0.72 & mean \textit{r}=0.51 & mean \textit{r}=0.57\\
 & s.d. \textit{r}=0.18 & s.d. \textit{r}=0.33 & s.d. \textit{r}=0.27\\
Power-law distribution & mean \textit{r}=0.70 & mean \textit{r}=0.46 & mean \textit{r}=0.60\\
 & s.d. \textit{r}=0.19 & s.d. \textit{r}=0.34 & s.d. \textit{r}=0.23\\
Uniform distribution & mean \textit{r}=0.68 & mean \textit{r}=0.49 & mean \textit{r}=0.60\\
 & s.d. \textit{r}=0.24 & s.d. \textit{r}=0.29 & s.d. \textit{r}=0.25\\
\hline
\end{tabular}
\end{table*}

\newpage
\begin{table*}[h]
\small
\caption{\textbf{Number of participants per election and number of participants with missing values.}}
\label{tab:tabS2}
\begin{tabular}{lllll}\hline
           \cline{1-5}
Election &	N complete sample & N non-voters & N missing values Democratic candidates)  & N missing values (Republican candidates)  \\ 
  \hline
1980 & 1,614 & 411 & 338 & 396\\
1984 & 2,257 & 539 & 583 & 435\\
1988 & 2,040 & 545 & 470 & 477\\
1992 & 2,485 & 562 & 567 & 358\\
1996 & 1,714* & 374 & 239 & 318\\
2000 & 1,807 & 376 & 440 & 493\\
2004 & 1,212 & 231 & 282 & 195\\
2008 & 2,322 & 509 & 368 & 390\\
2012 & 5,914 & 1,141 & 557 & 644\\
 \hline
\end{tabular}
\raggedright{\textit{Note}. *Of these 1,714 participants, 1,316 participants also participated during the election of 1992.}
\end{table*}

\begin{table*}[!htbp]
\caption{\textbf{Fit measures of the latent variable models.}}
\label{tab:tabS3}
\begin{tabular*}{1\linewidth}{@{\extracolsep{\fill}}lll}\hline
           \cline{1-3}
Candidate & One-factor model & Hierarchical model \\  \hline
Carter 1980 & $\chi(35)=1030.53$, $CFI=0.80$,  & $\chi(30)=528.58$, $CFI=0.90$,\\
 & $RMSEA=0.18$ & $RMSEA=0.14$\\
Reagan 1980 & $\chi(35)=411.40$, $CFI=0.91$, & $\chi(30)=198.74$, $CFI=0.96$,\\
 & $RMSEA=0.12$ & $RMSEA=0.08$*\\
Anderson 1980	& $\chi(35)=655.19$, $CFI=0.75$,  & $\chi(30)=290.38$, $CFI=0.90$,\\
  & $RMSEA=0.18$ & $RMSEA=0.12$*\\
 
 Mondale 1984 & $\chi(35)=1410.69$, $CFI=0.79$, & $\chi(31)=859.75$, $CFI=0.87$,\\
  &  $RMSEA=0.19$ & $RMSEA=0.15$\\
 
Reagan 1984 &	 $\chi(35)=1730.83$, $CFI=0.83$, & $\chi(31)=1194.73$, $CFI=0.88$,\\
  & $RMSEA=0.19$ & $RMSEA=0.17$*\\
 
Dukakis 1988 & $\chi(35)=1166.68$, $CFI=0.81$, & $\chi(31)=683.11$, $CFI=0.89$,\\
  & $RMSEA=0.18$ & $RMSEA=0.14$\\
 
Bush 1988 & $\chi(35)=940.98$, $CFI=0.87$, & $\chi(31)=593.40$, $CFI=0.92$,\\
 & $RMSEA=0.16$ & $RMSEA=0.13$*\\

Clinton 1992 & $\chi(35)=1536.91$, $CFI=0.83$, & $\chi(31)=995.36$, $CFI=0.89$,\\
 & $RMSEA=0.18$ & $RMSEA=0.15$*\\

Bush 1992 & $\chi(35)=1304.31$, $CFI=0.85$, & $\chi(31)=840.63$, $CFI=0.91$,\\
 &  $RMSEA=0.15$ & $RMSEA=0.13$*\\

Clinton 1996 & $\chi(35)=1492.00$, $CFI=0.83$, & $\chi(31)=908.92$, $CFI=0.90$,\\
 & $RMSEA=0.19$ & $RMSEA=0.16$*\\

Dole 1996 & $\chi(35)=1312.91$, $CFI=0.80$, & $\chi(31)=842.52$, $CFI=0.87$,\\
 & $RMSEA=0.19$ & $RMSEA=0.16$\\

Perot 1996 & $\chi(35)=428.94$, $CFI=0.78$, & Fit measures could not be computed*\\
 & $RMSEA=0.18$ & \\

Gore 2000 & $\chi(35)=1401.65$, $CFI=0.79$, & $\chi(31)=837.05$, $CFI=0.88$,\\
 & $RMSEA=0.19$ & $RMSEA=0.16$\\

Bush 2000 & $\chi(35)=1080.12$, $CFI=0.84$, & $\chi(31)=685.86$, $CFI=0.90$,\\
 &  $RMSEA=0.18$ & $RMSEA=0.15$\\

Kerry 2004 & $\chi(35)=771.78$, $CFI=0.86$, & $\chi(31)=458.36$, $CFI=0.92$,\\
 & $RMSEA=0.17$ & $RMSEA=0.14$*\\

Bush 2004 & $\chi(35)=1100.80$, $CFI=0.87$, & $\chi(31)=661.61$, $CFI=0.92$,\\
 &  $RMSEA=0.20$ & $RMSEA=0.16$*\\

Obama 2008 & $\chi(35)=1373.76$, $CFI=0.90$, & $\chi(31)=642.16$, $CFI=0.95$,\\
 & $RMSEA=0.16$ & $RMSEA=0.12$*\\

McCain 2008 & $\chi(35)=1553.26$, $CFI=0.86$, & $\chi(31)=863.92$, $CFI=0.92$,\\
 & $RMSEA=0.18$ & $RMSEA=0.14$\\

Obama 2012 & $\chi(35)=5863.21$, $CFI=0.91$, & $\chi(31)=2428.10$, $CFI=0.96$,\\
 & $RMSEA=0.20$ & $RMSEA=0.14$\\

Romney 2012 & $\chi(35)=5781.98$, $CFI=0.88$, & $\chi(31)=3021.35$, $CFI=0.94$,\\
 & $RMSEA=0.20$ &  $RMSEA=0.15$\\

 \hline
\end{tabular*}
\raggedright{\textit{Note}. *The covariance matrices of the latent variables of these models were not positive definitive. This indicates poor fit to the data.}
\end{table*}

\newpage
\subsection*{Supplementary Notes}
\subsubsection*{Supplementary Note 1: Analytical Solutions of the Hypotheses}
In the Ising model the intuition is that higher connection strength will allow for better prediction. Here we show this intuition is correct. 
\par\textbf{Logistic regression and the Ising model.}
The Ising model is part of the exponential family of distributions \cite{Brown1986, Young2005, Wainwright2008}. Let $G$ be a graph consisting of nodes in $V=\{1,2,\ldots,p\}$ and edges $(s,t)$ in $E\subseteq V\times V$. To each node $s\in V$ a random variable $X_{s}$ is associated with values in  $\{0,1\}$. The probability of each configuration $x$ depends on a main effect (external field) and pairwise interactions. It is sometimes referred to as the auto logistic-function \cite{Besag1974}, or a pairwise Markov random field, to emphasise that the parameter and sufficient statistic space are limited to pairwise interactions \cite{Wainwright2008}. Each $x_{s}\in \{0,1\}$ has conditional on all remaining variables (nodes) $X_{\backslash s}$ probability of success $\pi_{s}:=\Prob(X_{s}=1\mid x_{\backslash s})$.
The distribution for configuration $x$ of the Ising model is then 
\begin{align}
\Prob(x) = \frac{1}{Z(\theta)}\exp\left( \sum_{s\in V}m_{s}x_{s}+\sum_{(s,t)\in E}A_{st}x_{s}x_{t}\right)
\end{align}
which is clearly of the form of exponential family. In general, the normalisation $Z(\theta)$ is intractable, because the sum consists of $2^{p}$ possible configurations for $y\in\{0,1\}^{p}$; for example, for $p=30$ we obtain over 1 million configurations to evaluate in the sum in $Z(\theta)$ (see \citeNP{Wainwright2008} for lattice [Bethe] approximations).

The conditional distribution is again an Ising model \cite{Besag1974, Kolaczyk2009}
\begin{align}\label{eq:cond-prob}
\pi_{s}=\Prob(x_{s}=1\mid x_{\backslash s}) = \frac{\exp\left( m_{s}+\sum_{t:(s,t)\in E}A_{st}x_{t} \right)}{1+\exp\left( m_{s}+\sum_{t:(s,t)\in E}A_{st}x_{t} \right)}.
\end{align}
It immediately follows that the log-odds \cite{Besag1974} is 
\begin{align}\label{eq:log-odds}
\mu_{s}(x_{\backslash s})=\log\left(\frac{\pi_{s}}{1-\pi_{s}}\right)
=m_{s} +\sum_{t:(s,t)\in E}A_{st}x_{t}.
\end{align}
Note that the log-odds $\theta\mapsto\mu_{\theta}$ is a linear function, and so if $x=(1,x_{\backslash s})$ then $\mu_{\theta}=x^{\sf T}\theta$.

Recall that $\theta\mapsto \mu_{\theta}$ is the linear function $\mu_{\theta_{s}}(x_{\backslash s})= m_{s}+\sum_{t\in V\backslash s}A_{st}x_{t}$ of the conditional Ising model obtained from the log-odds (\ref{eq:log-odds}). Define $\mu_{s}:=\mu_{\theta_{s}}(x_{\backslash s})$. We use the notation that the node of interest $x_{i,s}$ is denoted by $y_{i}$ and we let the remaining variables and a 1 for the intercept be indicated by $x_{i}=(1,x_{i,\backslash s})$, basically leaving out the subscript $s$ to index the node, and only use it whenever circumstances demand it. Let the loss function be the negative log of the conditional probability $\pi$ in (\ref{eq:cond-prob}), known as a pseudo log-likelihood \cite{Besag1974}
\begin{align}\label{eq:psi-loss}
\psi(x,\mu) :=-\log \Prob (y\mid x) = -x\mu + \log(1+ \exp(\mu)).
\end{align}
\par\textbf{Monotonicity of prediction loss as a function of connectivity.}
In logistic regression there is a natural classifier that predicts whether $y_{i}$ is 1 or 0. We simply check whether the probability of a 1 is greater than 1/2, that is, whether $\pi_{i} > 1/2$. Because $\mu_{i}>0$ if and only if $\pi_{i}>1/2$ we obtain the natural classifier
\begin{align}
\mathcal{C}(y_{i}) = \mathbbm{1}\{ \mu_{i}>0 \}
\end{align}
where $\mathbbm{1}$ is the indicator function. This is 0-1 loss \cite{Hastie2015}. Sometimes the margin interpretation is used where the log of the conditional probability $\pi_{i,s}$ is used with variables in $\{-1,1\}$ (see \citeNP{Hastie2001}) . Let $z=2y-1$ such that for $x\in \{0,1\}$ we obtain $z\in \{-1,1\}$. The loss $\psi$ (pseudo log-likelihood) in (\ref{eq:psi-loss}) can then be rewritten as
\begin{align}\label{eq:log-loss}
\psi(z_{i},\mu_{i}) =\log(1+ \exp(-z_{i}\mu_{i})).
\end{align}
Often the logarithm with base 2 is chosen since then $\psi(z_{i},0)=1$. 
The classification translates to  
\begin{align}\label{eq:mis-loss}
\mathcal{C}(z_{i}) =\mathbbm{1}\{ z_{i}\mu_{i}>0 \}.
\end{align}
Logistic loss $\psi$ in (\ref{eq:log-loss}) is an upper bound to $\mathcal{C}$ in (\ref{eq:mis-loss}), and is 1 at the value of the margin $z_{i}\mu_{i}=0$, as shown in Supplementary Figure \ref{fig:figS1}. Here we use logistic loss $\psi$ as defined in (\ref{eq:psi-loss}) because it is more common. This function is strictly monotone decreasing. 
\par
It follows immediately from monotonicity that $\psi(z_{i},\mu)>\psi(z_{i},\mu^{*})$ if $\mu<\mu^{*}$. Of course, we have the same for the 0-1 loss: $\mathbbm{1}\{z_{i}\mu_{i}>0\}\ge \mathbbm{1}\{z_{i}\mu_{i}^{*}>0\}$ if $\mu<\mu^{*}$. 

If the average degree of each node were subtracted from the Hamiltonian $\mu_{s}$, then we obtain the Ising model without an external field. If we have
\begin{equation}
m_{s}=-\frac{1}{2}\sum_{t:(s,t)\in E}A_{st}
\end{equation}
then we see that
\begin{equation}
\mu_{s}=-\frac{1}{2}\sum_{t:(s,t)\in E}A_{st} + \sum_{t:(s,t)\in E}A_{st}x_{s}x_{t}=\sum_{t:(s,t)\in E}A_{st}(x_{s}x_{t}-1/2)
\end{equation}
 to the labeling $z=2y-1\in\{-1,1\}$, we obtain that the average is 0 (implying $m_{s}=0$), and so the Hamiltonian consists only of the interactions $\sum_{t:(s,t)\in E}A_{st}z_{s}z_{t}$. And so if $A_{st}\ge 0$ for all $s,t\in V$ then $\mu_{s}^{*}>\mu_{s}$ iff $A_{st}^{*} > A_{st}$, i.e., the Hamiltonian is larger if and only of the connectivity is larger. As seen above, this leads immediately to the monotonicity above. 
\par\textbf{Closeness and correlations}
Let $d_{st}(r)=\min\{1/r_{si}+1/r_{ij}+\cdots +1/r_{kt}: \forall i,j,k\in V\backslash \{s,t\}\}$ be the shortest distance in terms of Dijkstra's algorithm \cite{Wallis2007}, where $r_{ij}$ is the weight, in our case a (polychromic) correlation that are all positive. Then closeness is defined as
\begin{align}
c_{s}(r)=\left( \sum_{t\in V\backslash \{s\}} d_{st}(r) \right)^{-1}
\end{align}
The intuition is here that a node with high closeness will have connections or paths to other nodes with high correlations (weights). To see the intuition, consider node $s$ being connected only to node $t$ with correlation $r_{st}$. Then $c_{s}(r)=r_{st}$; if this correlation is high, then so is the closeness of this node. If there is more than one connection, we see that the shortest path $d_{st}(r)$ is low if all correlations $r_{ij}$ are high (close to $1$), implying that closeness $c_{s}(r)$ is high. 

Suppose we have two sets weights $R_{1}$ and $R_{2}$, inducing two graphs $G_{1}$ and $G_{2}$ with the same nodes and edge sets but with different weights. We pick a path between nodes $s$ and $t$, denoted by $P_{st}=\{(x_{0}=s,x_{1}),(x_{1},x_{2}),\ldots,(x_{k-1},x_{k}=t)\}$ of length $k$. Suppose that for this path we have 
\begin{equation}
\sum_{i=0}^{k} r_{1,i-1,i}\ge \sum_{i=0}^{k} r_{2,i-1,i}
\end{equation}
Then it follows that 
\begin{equation}
\sum_{i=0}^{k} \frac{1}{r_{1,i-1,i}}\le \sum_{i=0}^{k} \frac{1}{r_{2,i-1,i}}
\end{equation}
In other words, the higher the correlations the higher the closeness. This does not imply that any randomly drawn node connected to a node with high closeness will have a high correlation, only that on average the correlations will be higher if they are connected to a node with high closeness than if they are connected to a node with low closeness.

\subsubsection*{Supplementary Note 2: Alternative Analysis on Non-Voters}
	To investigate whether our results are robust to the inclusion of non-voters, we performed the same analyses but now including non-voters and labelling them as voters \textit{against} the focal candidates. The results of this analysis mirrored the results reported in the Results section: The correlation between connectivity and average impact remained high and significant (see Supplementary Figure \ref{fig:figS2}a). The same holds for the correlation between centrality and impact (see Supplementary Figure \ref{fig:figS2}b). The predicted impact remained very close to the actual impact (deviation median=0.06, deviation interquartile range=0.02-0.09) and outperformed both using the mean of all attitude elements (deviation median=0.10, deviation interquartile range=0.05-0.18, Wilcoxon-matched pairs test: \textit{V}=4006, \textit{P}$<$0.001, CLES=69.5\%) and using the means of the specific attitude elements (deviation median=0.08, deviation interquartile range=0.04-0.15, Wilcoxon-matched pairs test: \textit{V}=5670, \textit{P}$<$0.001, CLES=64.7\%).\par
	We also tested another prediction from our model regarding differences between voters and non-voters: That voters are expected to have a more densely connected network than non-voters. As can be seen in Supplementary Figure \ref{fig:figS2}c, attitude networks of voters were much more highly connected (mean=2.03, s.d.=0.43) than attitude networks of non-voters (mean=2.38, s.d.=0.34, Student's t-test: \textit{T}=2.86, \textit{P}$<$0.001, Cohen's \textit{D}=0.91).

\subsubsection*{Supplementary Note 3: Alternative Analysis on Independents}
	As our analysis are correlational, it is important to exclude the possibility that third variables affected the relations tested in this paper. The most likely variable to be such a confound is party identification. It is, for example, easy to imagine that party identification might affect the connectivity of attitude networks, the valence of the attitude, and for whom a person votes. We therefore reran our analyses including only participants, who do not identify with any political party. The results of this analysis mirrored the results reported in the Results section: The correlation between connectivity and average impact remained high and significant (see Supplementary Figure \ref{fig:figS3}a). The same holds for the correlation between centrality and impact (see Supplementary Figure \ref{fig:figS3}b). The predicted impact remained very close to the actual impact (deviation median=0.06, deviation interquartile range=0.03-0.11) and outperformed both using the mean of all attitude elements (deviation median=0.11, deviation interquartile range=0.05-0.17, Wilcoxon-matched pairs test: \textit{V}=4152, \textit{P}$<$0.001, CLES=66.2\%) and using the means of the specific attitude elements (deviation median=0.09, deviation interquartile range=0.04-0.16), Wilcoxon-matched pairs test: \textit{V}=5703, \textit{P}$<$0.001, CLES=62.7\%).

\subsubsection*{Supplementary Note 4: Alternative Analysis on Missing Values}
	To investigate whether our results are robust to imputation of missing values, we reran our analyses with imputing missing values using Predictive Mean Matching \cite{Little1988, vanBuuren2011}. The results of this analysis mirrored the results reported in the Results section: The correlation between connectivity and average impact remained high and significant (see Supplementary Figure \ref{fig:figS4}a). The same holds for the correlation between centrality and impact (see Supplementary Figure \ref{fig:figS4}b). The predicted impact remained very close to the actual impact (deviation median=0.06, deviation interquartile range=0.03-0.10) and outperformed both using the mean of all attitude elements (deviation median=0.12, deviation interquartile range=0.05-0.19, Wilcoxon-matched pairs test: \textit{V}=3724, \textit{P}$<$0.001, CLES=69.3\%) and using the means of the specific attitude elements (deviation median=0.10, deviation interquartile range=0.04-0.18, Wilcoxon-matched pairs test: \textit{V}=5337, \textit{P}$<$0.001, CLES=65.5\%).

\subsubsection*{Supplementary Note 5: Alternative Analysis on Networks Based on Different Numbers of Attitude Elements}
	To investigate whether our results are robust to our choice of attitude elements, we reran our analyses based on all available attitude elements (note that in this case the forecast analyses are not possible because for these analyses the same number of attitude elements for each election is necessary) and on the seven attitude elements that were assessed at each election. The results of these analyses mirrored the results reported in the Results section: For the analysis including all attitude elements, the correlation between connectivity and average impact remained high and significant (see Supplementary Figure \ref{fig:figS5}a). The same holds for the correlation between centrality and impact (see Supplementary Figure \ref{fig:figS5}b). \par
For the analysis including the seven attitude elements that were assessed at each election, the correlation between connectivity and average impact remained high and significant (see Supplementary Figure \ref{fig:figS5}c). The same holds for the correlation between centrality and impact (see Supplementary Figure \ref{fig:figS5}d). The predicted impact remained very close to the actual impact (deviation median=0.06, deviation interquartile range=0.03-0.09) and outperformed both using the mean of all attitude elements (deviation median=0.11, deviation interquartile range=0.06-0.18, Wilcoxon-matched pairs test: \textit{V}=1505, \textit{P}$<$0.001, CLES=68.7\%) and using the means of the specific attitude elements (deviation median=0.09, deviation interquartile range=0.05-0.16), Wilcoxon-matched pairs test: \textit{V}=2341, \textit{P}$<$0.001, CLES=65.2\%).

\subsubsection*{Supplementary Note 6: Alternative Analysis on Latent Variable Models}
	One might argue that the results reported in this article can also be expected when attitudes are conceptualized as latent variables and the responses on attitude elements are treated as indicators of the latent attitude. From this perspective, high (low) centrality of attitude elements would indicate high (low) factor loadings on the latent variable attitude and high (low) connectivity would indicate high (low) average factor loadings. In a purely statistical sense, this objection would be correct as factor loadings also reflect how much information a given attitude element holds on all other attitude elements. However, in our view the latent variable framework does not provide a sensible alternative for a data-generating model of the hypotheses put forward here. For such a model, one would have to assume that the latent variable attitude acts as common cause of the attitude elements. This assumption, however, is at odds with several key concepts in the attitude literature \cite{Dalege2016}, such as cognitive consistency \cite{Monroe2008}, ambivalence \cite{Thompson1995}, and the idea that attitudes are formed by attitude elements \cite{Fazio1995, Zanna1988}.\par
To further rule out that the latent variable framework provides an alternative explanation of our results, we investigated the fit of latent variable models on the data reported in this article. For each attitude toward each candidate at each election, we fitted two latent variable models. First, we fitted a one-factor model with all attitude elements loading on this single factor representing a latent attitude. Second, we fitted a hierarchical factor model with three or four first-order factors and one second-order factor representing a latent attitude. We fitted the hierarchical factor model because earlier research indicated that beliefs and feelings form different factors \cite{Breckler1984} and that negative and positive attitude elements form different factors \cite{vandenBerg2005}. In most of the data sets used here, no negative beliefs were assessed. For these data sets, we fitted a hierarchical factor model with beliefs, negative feelings, and positive feelings loading on different first-order factors, respectively. For the data sets in which negative beliefs were assessed, we fitted a hierarchical factor model with negative beliefs, positive beliefs, negative feelings, and positive feelings loading on different first-order factors, respectively. As can be seen in Supplementary Table \ref{tab:tabS3}, both the one-factor models and the hierarchical models fitted poorly. The latent variable framework thus appears to be an unlikely alternative explanation of our results. \par
This discussion on whether our results can also be explained by the latent variable framework is somewhat reminiscent of the discussion regarding the idea that instability of attitudinal responses is indicative of individuals holding nonattitudes \cite{Converse1970}. Several critiques of this idea pointed out that when measurement error is accounted for, individuals, who seemingly hold nonattitudes, show stable attitudes \cite{Achen1975, Ansolabehere2008, Judd1980}. A similar critique might apply to our findings. It is our view, however, that two findings speak against this critique. First, if we assume that the intercorrelations of attitude elements are determined only (or foremost) by measurement error, then the factor models we fitted should show good fit. This was clearly not the case. Second, systematic variation of intercorrelations would not be expected from the measurement error perspective. Thus, our finding that connectivity of attitudes correlates almost perfectly with the attitude's impact on behaviour would not be expected from the measurement error perspective.

\end{document}